\documentclass[journal,onecolumn]{IEEEtran}
\usepackage{amsmath}
\usepackage{amssymb}
\usepackage{graphicx}
\usepackage{citesort}
\usepackage{times}
\usepackage{subfigure}

\newcommand{\figref}[1]{Figure~\ref{#1}}

\newcommand{\complex}{\mathbb{C}}
\newcommand{\real}{\mathbb{R}}
\DeclareMathOperator{\tr}{tr}  %trace
\DeclareMathOperator{\diag}{diag}

\DeclareMathOperator{\trace}{tr}

\DeclareMathOperator{\adj}{adj}

\newcommand{\ones}{{\mathbf{1}}}
\newcommand{\nr}{{r}}
\newcommand{\nt}{{t}}
\newcommand{\st}{T}
\newcommand{\sr}{R}
\newcommand{\sti}{\tau}
\newcommand{\sri}{\rho}
\newcommand{\trimat}{\Gamma} %symbol for a triangular matrix
\newcommand{\trimati}{d} %entry of triangular matrix.

\newcommand{\expect}[1]{{\mathcal{E}}_{#1}\!}
\newcommand{\herm}[1]{{#1}^\dag}
   
\renewcommand{\expect}[2][]{{\mathrm{E}_{#1}\!\left\{ #2 \right\}}}

\newcommand{\wishart}[3]{W_{#1}\left(#2,#3\right)}
\newcommand{\normal}[4]{{\mathcal{N}}_{#1,#2}\left(#3,#4\right)}
\newcommand{\psifunc}[2]{{\Psi\!\left(#1, p(#2)\right)}}
\newcommand{\psifuncrand}[2]{{\psi\!\left(#1, #2\right)}}
\newcommand{\snr}{\gamma}
\newcommand{\symboltime}{k}
\newcommand{\norm}[1]{\left\|#1\right\|}

\newtheorem{lemma}{Lemma}
\newtheorem{theorem}{Theorem}

\newtheorem{example}{Example}

\newtheorem{problem}{Problem}
\newtheorem{algorithm}{Algorithm}

\newcommand{\figwidth}{75mm}%{0.8\columnwidth}

\begin{document}

%%%%%%%%%%%%%%%%%%%%%%%%%%%%%%%%%%%%%%%%%%%%%%%%%%%%%%%%%%%%

\title{Optimal Transmit Covariance for Ergodic MIMO Channels}

\author{Leif~W.~Hanlen,~%\IEEEmembership{Member,~IEEE,} %
Alex~J.~Grant,~%\IEEEmembership{Member,~IEEE,}%
\thanks{L. Hanlen is with National ICT Australia, Canberra, Australia, email: {\tt leif.hanlen@nicta.com.au} National ICT Australia  is funded through the Australian Government's \emph{Backing Australia's Ability initiative}, in part through the Australian Research Council.}%
\thanks{A. Grant is with the Institute for Telecommunications Research, University of South Australia, Australia, email: {\tt alex.grant@unisa.edu.au}}%
\thanks{A part of this work appeared in \cite{HanGrant05},\cite{HanGra:0205},\cite{Grant:0205}}
}

\maketitle

\begin{abstract} 
In this paper we consider the computation of channel capacity for
ergodic multiple-input multiple-output channels with additive white
Gaussian noise. Two scenarios are considered. Firstly, a time-varying
channel is considered in which both the transmitter and the receiver
have knowledge of the channel realization. The optimal transmission
strategy is water-filling over space and time. It is shown that this
may be achieved in a causal, indeed instantaneous fashion. In the
second scenario, only the receiver has perfect knowledge of the
channel realization, while the transmitter has knowledge of the
channel gain probability law. In this case we determine an optimality
condition on the input covariance for ergodic Gaussian vector channels
with arbitrary channel distribution under the condition that the
channel gains are independent of the transmit signal. Using this
optimality condition, we find an iterative algorithm for numerical
computation of optimal input covariance matrices. Applications to
correlated Rayleigh and Ricean channels are given. 
\end{abstract}

\section{Introduction}
Shannon theoretic results for multiple-input multiple-output (MIMO) fading
channels~\cite{Telatar1199,Gans98} have stimulated a large amount of research
activity, both in the design of practical coding strategies and in
extension of the theory itself. 

From an information theoretic point of view, the main problem is to
find the maximum possible rate of reliable transmission over
$\nt$-input, $\nr$-output additive white Gaussian noise channels of
the form
\begin{equation} \label{E:1.the_channel}
  y[\symboltime] = \sqrt{\snr} H[\symboltime] x[\symboltime] + z[\symboltime]
\end{equation}
where $y[\symboltime]\in\complex^{\nr\times1}$ is a complex column
vector of matched filter outputs at symbol time
$\symboltime=1,2,\dots,N$ and
$H[\symboltime]\in\complex^{\nr\times\nt}$ is the corresponding matrix
of complex channel coefficients. The element at row $i$ and column $j$
of $H[\symboltime]$ is the complex channel coefficient from transmit
element $j$ to receive element $i$. The vector
$x[\symboltime]\in\complex^{\nt\times 1}$ is the vector of complex
baseband input signals, and $z[\symboltime]\in\complex^{\nr\times 1}$
is a complex, circularly symmetric Gaussian vector with
$\expect{n[\symboltime] \herm{n[\symboltime]}} = I_{\nr}$.  The
superscript $\herm{(\cdot)}$ means Hermitian adjoint and $I_\nr$ is
the $\nr\times\nr$ identity matrix. Let $n=\max(\nt,\nr)$ and
$m=\min(\nt,\nr)$.

Transmission occurs in codeword blocks of length $N$ symbols. 
Let $x_N\in\complex^{\nt}$ and $y_N\in\complex^{\nr}$ be the column
vectors resulting from stacking $x[1],x[2],\dots,x[N]$ resp.
$y[1],y[2],\dots,y[N]$. Further let $H_N$ be the block-diagonal matrix
with diagonal blocks $H[\symboltime]$.

A transmitter power constraint 
\begin{equation}\label{eq:PowerConstraint}
  \frac{1}{N} \norm{x_N}_2^2 \leq 1
\end{equation}
is enforced, where $N$ is the codeword block length. This power
constraint has been explicitly written out this way to remind the
reader that power constraints such as this, commonly written
$E[\norm{x[\symboltime]}_2^2]\leq 1$ are long-term \emph{average}
power constraints, not deterministic per-symbol, or per-input
constraints, see~\cite[p. 329]{Cover91}. Accordingly, the
signal-to-noise ratio is defined as $\snr$. The covariance matrix of
input sequences of length $N$ is defined as the $N\nt \times N\nt$
matrix
\begin{equation}
Q_N = \expect{x_N\herm{x_N}}
\end{equation}
and hence the power constraint can also be written as $\tr(Q_N)\leq N$. Also define the per-symbol input covariance matrices $Q[\symboltime]=\expect{x[\symboltime]\herm{x[\symboltime]}}$, which appear as principal sub-matrices in $Q_N$. In the case of memoryless transmission, $Q_N$ is a block diagonal matrix with diagonal blocks $Q[\symboltime]$.

The power constraint~\eqref{eq:PowerConstraint} assumes that the power
received from the collection of transmit signals at any point in space
(e.g.. at some imaginary point close to the transmitter) is given by the
summation of the individual signal powers, ie. zero mutual coupling.

There are several possibilities for the amount of side information
that the receiver or transmitter may possess regarding the channel
process $H[\symboltime]$. \emph{Perfect} side information shall mean
knowledge of the \emph{realizations} $H[\symboltime]$, while
\emph{statistical} side information refers to knowledge of the
\emph{distribution} from which the $H[\symboltime]$ are selected. Perfect receiver side information will be assumed throughout the
paper.

There are several categories of channels~(\ref{E:1.the_channel}) that
have been investigated in the literature:
\begin{enumerate}
\item\label{deterministic} Channels in which $H[\symboltime]$,
  is a given sequence of channel matrices, known to both the transmitter and receiver.
\item\label{ergodic} Ergodic channels in which the $H[\symboltime]$,
  $\symboltime=1, 2, \dots$ are random matrices, selected
  independently of each other and independently of the $x[\symboltime]$,
  according to some matrix probability density function $p_H$, which is known at the transmitter. The specific channel realizations are unknown at the transmitter, but are known at the receiver.
\end{enumerate}
Under the assumption of additive Gaussian noise and perfect receiver side information, the optimal
input distribution is Gaussian, and the main problem is therefore the
determination of the capacity achieving input covariance matrix $Q_N$.

For a given input covariance, the information rate for case \ref{deterministic} is (adopting a modification of the notation of~\cite{Telatar1199}),
\begin{equation}
  \psi(Q_N,H_N) = \frac{1}{N} 
  \log\det\left(I_{N\nt} + H_N Q_N \herm{H_N}\right). 
  \label{E:def-psi-rand}
\end{equation}
The capacity is found by maximizing the information rate.
\begin{problem}[Gallager~\cite{Gallager68}]\label{Prob:1}
\begin{align*}
\max_{Q_N} &\, \psi(Q_N,H_N)
\intertext{subject to}
\frac{1}{N}\tr(Q_N) &\leq 1 \\
Q_N      &\geq    0
\end{align*}
\end{problem}
Note that since $\psi$ is a function of $H_N$, the optimal covariance matrix will in general be a function of $H_N$.

Telatar~\cite{Telatar1199} obtained the solution of
Problem~\ref{Prob:1}, when $H[\symboltime]=H$ for all
$\symboltime=1,2,\dots$. Following Gallager~\cite{Gallager68}, the
solution is obtained by solution of the Kuhn-Tucker conditions, and
results in a water-filling interpretation,
\begin{align}
  C &= \sum_{i:\lambda_i^{-1}\leq\mu} \log \mu \lambda_i,\quad
  \text{where $\mu$ is such that}
  \label{48831eq:capacityC} \\
  \snr &= \sum_{i:\lambda_i^{-1}\leq\mu} \mu - \lambda_i^{-1} 
  \label{48831eq:capacityP}
\end{align}
and $\lambda_i$, $i=1,2,\dots,m$ are the non-zero eigenvalues of
$H\herm{H}$. The optimal transmit covariance matrix is independent of $\symboltime$ and is given by $Q[\symboltime] = Q = \herm{V} \Gamma
V$, where $V$ is the matrix of right singular vectors of $H$ and
$\Gamma=\diag\{\max(0,\mu-1/\lambda_i)\}$. 

The information rate in the ergodic case is $\Psi = \expect{\psi}$ and subject to the assumptions in case \ref{ergodic} above, reduces to a symbol-wise expectation with respect to $p_H$,
\begin{equation}
 \Psi(Q,p_H) = \expect{\log\det\left(I + HQ\herm{H}\right)} \label{E:def-psi}
\end{equation}
where $Q = Q[\symboltime]$ is $\nt\times\nt$ covariance matrix for each symbol.  In this case, capacity is found via solution of
\begin{problem}[Telatar~\cite{Telatar1199}]
  \label{Prob:1a}
\begin{align*}
\max_{Q} &\, \Psi(Q, p_H)
\intertext{subject to}
\tr(Q) &\leq 1 \\
Q      &>    0
\end{align*}
\end{problem}
Since $\Psi$ is an expectation with respect to $p_H$, the optimal $Q$ will depend on $p_H$, rather than the realizations $H[\symboltime]$. 

One common choice for $p_H$ is a Gaussian density. We will use the
notation $\normal{\nt}{\nr}{M}{\Sigma}$ to mean a Gaussian density
with $\nr\times\nt$ mean matrix $M$ and $\nr\nt\times\nr\nt$
covariance matrix $\Sigma = \expect{h\herm{h}}$ where $h$ is formed by
stacking the columns of the matrix into a single vector. This allows
for arbitrary correlation between elements. Common special cases
include i.i.d. unit variance entries, $\normal{\nt}{\nr}{0}{I}$
(corresponding to independent Rayleigh fading) and the so-called
Kronecker correlation model $\normal{\nt}{\nr}{M}{R\otimes T}$. The
latter model corresponds to separable transmit $T$ and receive
correlation $R$, and may be generated via $M + R^{1/2} G T^{1/2}$
where $G\sim\normal{\nt}{\nr}{0}{I}$.  For
$H[\symboltime]\sim\normal{\nt}{\nr}{0}{I}$ Telatar showed that the
optimizing $Q=I_\nt/\nt$, meaning that it is optimal to transmit
independently with equal power from each antenna. Thus in that case
\begin{equation}\label{E:1.1}
  C = \expect{\log\det\left(I_\nr +\frac{\snr}{\nt} H\herm{H}\right)}. 
\end{equation}
Telatar also gave an expression for computation of~\eqref{E:1.1}, and
several other expressions have subsequently been found
\cite{Gra02,BigTar04,HanGra:04IT-unpub}.

Finally, Telatar considered a variation on case \ref{deterministic},
with time-invariant $H[\symboltime]=H$ and perfect receiver side
information, but only statistical transmitter side information. This
requires the notion of outage probability. It was conjectured that the
optimal transmission strategy, minimizing the outage probability, is
equal power signals from a subset of antennas. We do not consider
outage probability in this paper.

It is clear from these results that the degree of channel knowledge at
the transmitter has a significant effect on the optimal transmission
strategy.

Extensions to the theory have taken several directions, for example
extending the ergodic capacity results to channel matrices whose
elements are no longer independent of each other.  ``One-ring''
scatterer models, resulting in single-ended correlation structure
$H\sim\normal{\nt}{\nr}{0}{I\otimes T}$ were considered
in~\cite{ShiFos00TC}. Bounds on capacity were obtained in that work,
assuming $Q=I/\nt$. Subsequently, a series of papers appeared,
adopting the same single-ended correlation model. In
\cite{JafVishGold:ICC01} it was shown that for
$H\sim\normal{\nt}{\nr}{0}{I\otimes T}$ it is optimal to transmit
independently on the eigenvectors of $T$. Majorization results were
obtained showing that stronger modes should be allocated stronger
powers, and optimal $Q$ were found using numerical optimizations. No
conditions for optimality were given. In \cite{ChiWinZan03IT}, a
closed-form solution for the characteristic function of the mutual
information assuming $Q=I/\nt$ was found for the same single-ended
correlation model. In \cite{SimMouJSAC03}, the special case of $\nt=2$
was considered, where optimization of $Q$ could be performed, once
again assuming no receiver correlation, $R=I$.

Asymptotic large systems ($\nr,\nt\to\infty$ with $\nr/\nt\to$ a
constant) capacity results have been obtained in \cite{Tse0302}, for
the more general case $H\sim\normal{\nt}{\nr}{0}{R\otimes T}$, but
under the assumption $Q=I/\nt$. Asymptotic results for arbitrary $Q$ were
considered in \cite{MarOtt04}, where the asymptotic distribution of
the mutual information was found to be normal. Large-systems results
have been obtained in \cite{TulLozVerdu0705}, concentrating on the
case where the eigenvectors of the optimal $Q$ can be identified by
inspection.

Closed form solutions have been obtained for the mutual information of
single-ended correlated channels
\cite{AlfTulLozVer:ISSSTA04,HanGra:04IT-unpub} and for
$H\sim\normal{\nt}{\nr}{0}{R\otimes T}$, \cite{KieSpe04IZS,SimMou04}.

Non-zero mean multiple-input, single-output channels were considered
in \cite{VisMad01,MousSim:IT03}.  In those papers, results were
obtained for non-zero mean, in the absence of transmitter correlation,
and for non-trivial transmitter correlation, with zero mean. Further
results for non-zero mean channels have been presented in
\cite{KimLap03}, which reports some majorization results on mutual
information, with respect to the eigenvalues of the mean matrix. Exact
distributions of mutual information have been obtained in for $\nt=2$
or $\nr=2$. Asymptotic expressions for the mutual information have
been presented in \cite{CotDeb04}, for arbitrary $Q$, and non-central,
uncorrelated fading.

Other
researchers~\cite{Marzetta0199,Goldsmith1197,Medard2000,ZheTse02IT}
have examined variations on the amount of information available at
transmitter and receiver.

Previous work such
as~\cite{Gallager68,JafVishGold:ICC01,Telatar1199,MousSim:IT03,SimMouJSAC03,TulLozVerdu0705}
on Gaussian vector channels focused on cases when the eigenvectors of
the optimal input covariance can be easily determined by inspection of
the channel statistics, and the problem becomes one of optimizing the
eigenvalues of the input covariance. This approach does not lend
itself to arbitrary non-deterministic channels: for example where the
channel mean and covariance are not jointly diagonalizable or where
the probability density is not in Kronecker
form~\cite{OzcHer03EL,Poll04}.

This paper provides general solutions of Problems \ref{Prob:1} and
\ref{Prob:1a}. The latter provides a solution to \cite[open problem 1 and
2]{GolJaf03JSAC}, albeit not in closed form.

 In Section~\ref{sec:TxFullCSI} we extend the
water-filling result to ergodic channels where the transmitter has
perfect knowledge of the channel realization $H[\symboltime]$ at each
symbol. In Section~\ref{sec:TxStatisticalCSI} we relax the degree of
transmitter channel knowledge and consider the ergodic channel with
arbitrary channel distribution $p_H$, such that $p_H$, but not
$H[\symboltime]$ is known to the transmitter.

The semidefinite constraint $Q\geq 0$ in Problem \ref{Prob:1a} would
normally make the optimization difficult. However, in several cases,
the eigenvectors of the optimal $Q$ may be identified a-priori, which
reduces the problem to an optimization over the space of probability
vectors. In independent work,~\cite{TulLozVerdu0705} has found similar
results to those presented in this paper for this "diagonalizable"
case. We avoid the requirement of diagonalizing $Q$. Our main result
is the determination of the capacity achieving covariance for
arbitrary ergodic channels. This is achieved by finding necessary and
sufficient conditions for optimality, which in turn yield an iterative
procedure for numerical optimization of $Q$, which finds the optimal
eigenvectors in addition to the optimal eigenvalues.
In each section we provide numerical examples that illustrate the application of the main results. Conclusions are drawn
in Section~\ref{sec:Conclusion}. All proofs are to be found in the
Appendix.

\section{Perfect Transmitter Side  Information}\label{sec:TxFullCSI}
As described above, Telatar \cite{Telatar1199} solved Problem
\ref{Prob:1} for time-invariant deterministic channels. There are
cases of interest however when the transmitter and receiver have
perfect side information, but the channel is time-varying. One model
for this case is to suppose that $H[\symboltime]$ is indeed
time-varying, and that this sequence is a realization of a random
process, in which each $H[\symboltime]$ is selected independently at
each symbol $\symboltime$ (and independently of the $x[\symboltime]$)
according to some probability law $p_H$, so the channel remains
memoryless.

Subject to this model, we seek a solution to Problem \ref{Prob:1}, in
which the sequence of channel matrices are generated i.i.d. according
to $p_H$.  It is tempting to simply average \eqref{48831eq:capacityC}
over the ordered eigenvalue density,  $p_\Lambda(\lambda_1,\dots,\lambda_m)$,
associated with $p_H$ (see for example
\cite{TulLozVer:ISSSTA04}),
\begin{equation}\label{48831eq:wfaverage}
\expect{C} =
  \int p(\lambda_1,\dots,\lambda_m) \sum_{i:\lambda_i^{-1}\leq\mu}
  \log \mu \lambda_i \, d\Lambda
\end{equation}
This quantity is however in general not the capacity of the
channel~(\ref{E:1.the_channel}) with $H[\symboltime]\sim p_H$. A
simple counter-example suffices to show the problem.

\begin{example}
  Consider a single-input single-output channel, $r=t=1$, and let $p_H(\epsilon)=p_H(1)=1/2$ where $\epsilon>0$.
  Then according to~(\ref{48831eq:wfaverage}) in which water-filling
  precedes averaging, the resulting information rate is
  $\log\left(1+\snr\epsilon^2\right)/4 + \log(1+\snr)/4$ which as
  $\epsilon\rightarrow 0$ approaches $\log(1+\snr)/4$.

  It is obvious however that as $\epsilon\rightarrow 0$, the transmitter
  should only transmit in symbol intervals in which $H=1$, resulting in
  the capacity $\log(1+\snr)/2$ which is a factor of two greater than the
  previous approach.
\end{example}
The problem with~(\ref{48831eq:wfaverage}) is that it precludes
optimization of the transmit density over \emph{time} as well as
space. The rate \eqref{48831eq:wfaverage} is maximal only under the
assumption of a short-term power constraint $\tr(Q[\symboltime])=1$,
rather than the long-term constraint $\tr(Q_N)=N$.

The following Theorem, is proved by solving
the input distribution optimization problem from first principles (see
Appendix).
\begin{theorem}\label{48831th:capacity}
  Suppose that the channel matrices $H[\symboltime]$ of an ergodic
  MIMO channel~\eqref{E:1.the_channel} are selected i.i.d. each symbol
  $\symboltime$ according to a matrix density $p_H$ which possesses an
  eigenvalue density $f_\lambda$. The capacity of this
  channel with prefect channel knowledge at both the transmitter and
  the receiver is given by
  \begin{align}
    \frac{C}{m} &= \int_{\xi^{-1}}^{\infty}
    \log\left(\xi\lambda\right)\,
    f_\lambda(\lambda)\,d\lambda \quad \text{where $\xi$ is such
    that}\label{48831eq:capacityC3}\\ 
    \frac{\snr}{m} &= \int_{\xi^{-1}}^{\infty} \left(\xi -
      \frac{1}{\lambda}\right)\, f_\lambda(\lambda)\,d\lambda.
    \label{48831eq:capacityP3}
  \end{align}
\end{theorem}
It is interesting to note that not only does this Theorem yield the
actual capacity, as opposed to the rate given
by~\eqref{48831eq:wfaverage}, it is also easier to compute in most
cases, since it is based on the distribution of an unordered
eigenvalue.

Water-filling over space and time has been addressed to a limited extent in the
literature. Tse and Viswanath give the result, without proof~\cite[Section 8.2.34]{TseVis04}. Goldsmith
also writes down the optimization problem (without solution) in~\cite[Equation (10.16)]{Gol04}, and also in~\cite{Goldsmith1197}. The
correct space-time water filling approach is also implicit
in~\cite{JayPoo03IT}, although no proof or discussion is offered. 

Let us now examine the optimal transmit strategy in more detail. Let
$H[\symboltime] = U[\symboltime] \Lambda[\symboltime] V[\symboltime]$ be the singular value decomposition of
$H[\symboltime]$ and let $H_N$, $U_N$, $V_N$ and $\Lambda_N$ be the corresponding
block diagonal matrices. Then the
singular value decomposition of the block diagonal matrix $H_N$ is
\begin{equation}\label{48831eq:svd}
  H_N = U_N \Lambda_N V_N.
\end{equation}
This follows directly from the block-diagonal structure of $H_N$.
The fact that the singular vectors are also in block-diagonal form is
important from an implementation point of view. If it had turned out
that $H_N$ had full singular vector matrices, the optimal
transmission strategy would be non-causal. 

The optimal transmit strategy uses a block-diagonal input
covariance matrix,
\begin{equation}
  Q_N = \diag\left\{\herm{V}[1] \Gamma[1] V[1], \dots, \herm{V}[N] \Gamma[N]
  V[N] \right\} 
\end{equation}
where $\Gamma[\symboltime] = \left(\xi I -
  \left(\Lambda[\symboltime]\right)^{-1}\right)^+$, using the notation $(\cdot)^+$ which
replaces any negative elements with zero.  The block-diagonal
structure means that the input symbols are correlated only over space,
and not over time.  At time $\symboltime$, the input covariance is
$Q[\symboltime] =
\herm{V}[\symboltime]\Gamma[\symboltime]V[\symboltime]$. Thus the
optimal transmit strategy is not only causal, but is instantaneous,
i.e. memoryless over time. At time $\symboltime$, the transmitter does
not need to know any past or future values of $H[j]$, $j>i$ and $j<i$
in order to construct the optimal covariance matrix.

The key thing to note from Theorem~\ref{48831th:capacity} is that the
required power allocation is still water-filling on the eigenvalues of
$H[\symboltime]\herm{H}[\symboltime]$, but that the water level $\xi$ is chosen to satisfy the
actual average power constraint, rather than a symbol-wise power
constraint. At any particular symbol time, the transmitter uses a
power allocation $(\xi-1/\lambda)^+$ for each eigenvalue $\lambda$ of
$H[\symboltime]\herm{H}[\symboltime]$, noting that $\xi$ is selected according
to~(\ref{48831eq:capacityP3}) rather than on a per-symbol
basis,~(\ref{48831eq:capacityP}). This does not require any more
computation that symbol-wise water filling. In fact, it is simpler,
since the transmitter only needs to compute the water level $\xi$
once. Not only does space-time water filling give a higher rate, it is
in this sense easier to implement.

One possible argument against the use of space-time water-filling is
that with this approach, there is a variable amount of energy
transmitted at each symbol interval. In some cases that would
certainly be undesirable (such as systems using constant envelope
modulation). 
\begin{theorem}\label{48831th:papr}
  The peak-to-average power ratio resulting from space-time
  water-filling,~(\ref{48831eq:capacityC3}),~(\ref{48831eq:capacityP3}) on an
  ergodic channel with average power constraint $\snr$ and unordered
  eigenvalue density $f(\lambda)$ such that $E[1/\lambda]$ exists
  is upper-bounded
\begin{equation*}
  \text{PAPR} \leq 1 + \frac{m}{\snr}\, E\left[\lambda^{-1}\right].
\end{equation*}
\end{theorem}
This is a particularly simple characterization of the PAPR. The term
$m E[1/\lambda]/\snr$ is the ratio of the average inverse eigenvalue to
the average symbol energy per eigen-mode.

It is also straightforward to compute the information rate $I$ that
results from adjusting the space-time water-filling solution to
accommodate a peak-power limitation $\snr_{\max}$,
\begin{align*}
  \frac{I}{m} &= \int_{\xi^{-1}}^{(\xi-\snr_{\max})^{-1}}
  \log\left(\xi\lambda\right)\,
  f(\lambda)\,d\lambda \quad \text{where $\xi$ is such
    that}\\ 
  \frac{\snr}{m} &= \int_{\xi^{-1}}^{(\xi-\snr_{\max})^{-1}} \left(\xi -
    \frac{1}{\lambda}\right)\, f(\lambda)\,d\lambda.
\end{align*}
Note that this is not the same as the capacity of the peak-power
constrained channel. In practice however, it may be of interest, since
powers approaching $\xi$ are typically transmitted with vanishing
probability. It is therefore of interest to consider the probability
density function $q(\gamma)$ of the per-eigenvector transmit power,
$\gamma = \xi-1/\lambda$. The obvious transformation yields the
density function.
\begin{theorem}
  The probability density function $q(\gamma)$ of the energy
  $\gamma=\xi-1/\lambda$ transmitted on each eigenvector according
  to~(\ref{48831eq:capacityC3}),~(\ref{48831eq:capacityP3}) is given by
  \begin{equation*}
  q(\gamma) = F\left(\xi^{-1}\right)\delta(\gamma) + \frac{f\left(\left(\xi -
  \gamma\right)^{-1}\right)}{ 
  \left(\xi-\gamma\right)^{2}},
\end{equation*} 
where $f(\cdot)$ is the unordered eigenvalue density, $F(\cdot)$ is
the corresponding cumulative distribution and $\delta$ is the Dirac
delta function. The point mass at $\gamma=0$ corresponds to the
probability of transmitting nothing on that channel (when the gain is
less than $1/\xi$).
\end{theorem}

%% *** This next part needs some more work ***
%% *** e.g. Compute limit at P->0          ***
%% Obtaining precise knowledge of the channel realization at the
%% transmitter may be infeasible in certain circumstances. How much extra
%% rate is gained by this space-time water-filling method over the case
%% where the transmitter does not know the channel realization?
%% The rate difference is, 
%% \begin{multline}
%%   m\int_{\xi^{-1}}^\infty \log(\lambda\xi)p(\lambda)\,d\lambda - m\int_0^\infty
%%   \log(1+P\lambda) p(\lambda) \, d\lambda \\
%% = m E\left[ \log\left(\frac{\lambda\xi}{1+P\lambda}\right)\right] -
%%   m \int_{0}^{\xi^{-1}} \log(\lambda\xi) p(\lambda)\, d\lambda 
%% \end{multline}
%% which is asymptotic to $m\log(\xi/P)$ as $P\rightarrow\infty$. Of
%% course $\xi$ is a function of $P$ and as $P\rightarrow\infty$,
%% $\xi\rightarrow P$.

The following examples show some simple applications of the preceding
space-time water-filling result.
\begin{example}[Parallel On-Off Channel]
  Consider an $m$-input, $m$-output channel with eigenvalue density
  $(1-p)\delta(\lambda) + p\delta(\lambda-1)$. There are $m$ parallel
  channels and each channel is an independent Bernoulli random
  variable. With probability $p$, a channel is ``on'' and with
  probability $1-p$ it is ``off''. 
  
  Spatial water-filling yields the rate
  \begin{equation*}
    E\left[\frac{k}{2} \log\left(1 + \frac{P}{k}\right)\right],
  \end{equation*}
  where $k\sim\text{Binomial}(m,p)$. It is straightforward to show
  however that the capacity is
  \begin{align*}
    C &= \frac{E[k]}{2}\log\left(1 + \frac{P}{E[k]}\right) \\
    &=  \frac{mp}{2}\log\left(1 + \frac{P}{mp}\right).
  \end{align*}
which, as expected is strictly larger than the former rate, a fact
that can be seen from Jensen's inequality. 
\end{example}

\begin{figure}[tbp]
  \centering\setlength{\unitlength}{1mm}
  \begin{picture}(80,50)
    \put(0,0){\includegraphics*[width=80mm]{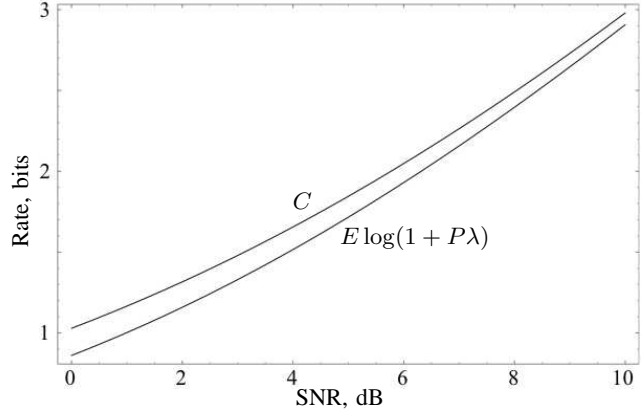}}
    \put(40,0){\makebox(0,0)[t]{\small SNR, dB}}
    \put(-1,25){\makebox(0,0)[r]{\rotatebox{90}{\small Rate, bits}}}
    \put(35,25){\makebox(0,0){{\small $C$}}}
    \put(40,20){\makebox(0,0)[l]{{\small $E\log(1+P\lambda)$}}}
  \end{picture}
  \caption{Single-input, single-output Rayleigh channel.}
  \label{48831fig:g1}
\end{figure}

\begin{example}[Rayleigh, $t=r=1$]
  Consider the single-input, single-output Rayleigh fading
  channel. Then $f(\lambda) = e^{-\lambda}$ and $\xi$ is the solution
  to 
  \begin{equation*}
    \xi e^{-1/\xi} + \Gamma\left(0,\xi^{-1}\right) = P,
  \end{equation*}
  where $\Gamma(a,x)$ is the incomplete Gamma
  function~\cite[(8.350.2)]{Gradshteyn00}.  \figref{48831fig:g1} compares the
  resulting capacity to the rate obtained via per-symbol
  water-filling. Note that in this case, the latter corresponds to the
  capacity when the transmitter does not know the channel realization.
  In other words, application of the incorrect method results in
  ignoring the channel knowledge at the receiver.
\end{example}

\begin{figure}[tbp]
  \centering\setlength{\unitlength}{1mm}
  \begin{picture}(80,50)
    \put(0,0){\includegraphics*[width=80mm]{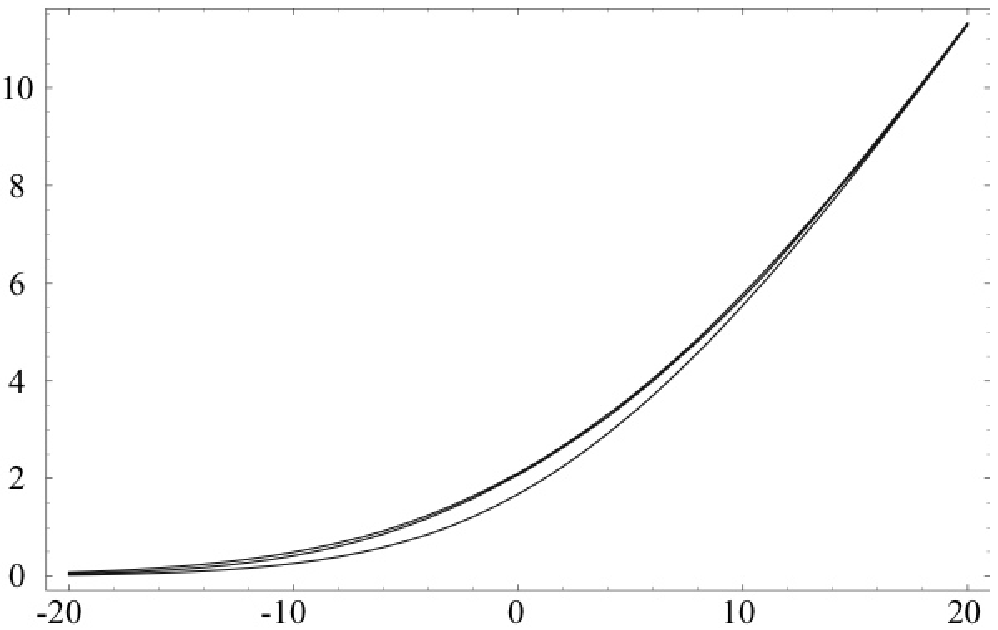}}
    \put(40,0){\makebox(0,0)[t]{\small SNR, dB}}
    \put(-1,25){\makebox(0,0)[r]{\rotatebox{90}{\small Rate, bits}}}
    \put(54,25){\makebox(0,0){{\small $C$}}}
    \put(56,20){\makebox(0,0)[l]{{\small $E\log(1+P\lambda)$}}}
  \end{picture}
  \caption{Rayleigh channel, $t=r=2$.}
  \label{48831fig:gabs}
\end{figure}

\begin{figure}[htbp]
  \centering\setlength{\unitlength}{1mm}
  \begin{picture}(80,50)
    \put(0,0){\includegraphics*[width=80mm]{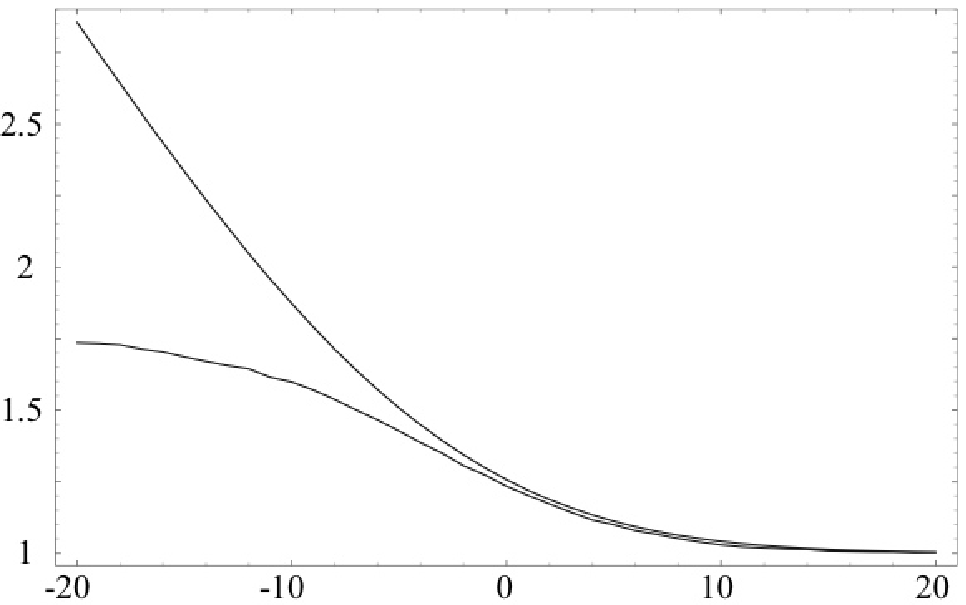}}
    \put(40,0){\makebox(0,0)[t]{\small SNR, dB}}
    \put(-1,25){\makebox(0,0)[r]{\rotatebox{90}{\small Relative
    capacity gain}}}
    \put(30,32){\makebox(0,0){{\small Space-Time}}}
    \put(15,18){\makebox(0,0){{\small Space}}}
  \end{picture}
  \caption{Rayleigh channel, $t=r=2$.}
  \label{48831fig:grel}
\end{figure}

\begin{example}[Rayleigh $t=r=2$]
  Consider the two-input, two-output Rayleigh fading
  channel. Then $f(\lambda) = \frac{2 + \left( \lambda - 2 \right)
  \,\lambda }{2\,e^{\lambda }}$ and $\xi$ is the solution to
\begin{equation*} {{e }^{-1/\xi }} 
    (2 \xi +1)- 2\Gamma \left(0,\xi^{-1}\right) = P.
\end{equation*}
\figref{48831fig:gabs} compares the resulting capacity to the rate
obtained via per-symbol water-filling and to the rate obtained with
$Q=P I_t$. The curves for space-time water-filling and spatial
water-filling almost coincide on this figure. This is however hiding
the additional gain provided by space-time water-filling at low SNR.
\figref{48831fig:grel} shows the relative gains, compared to $Q=PI_t$
for space-only and space-time water-filling. Obviously, as
$SNR\rightarrow\infty$, both gains approach 1, since there is
asymptotically no benefit in water filling of any kind.  At SNR below
$0$ dB, space-time water-filling yields significant benefit compared
to water-filling only over space.
\end{example}

%\fi

\begin{figure}[htbp]
  \centering\setlength{\unitlength}{1mm}
  \begin{picture}(80,50)
    \put(0,0){\includegraphics*[width=80mm]{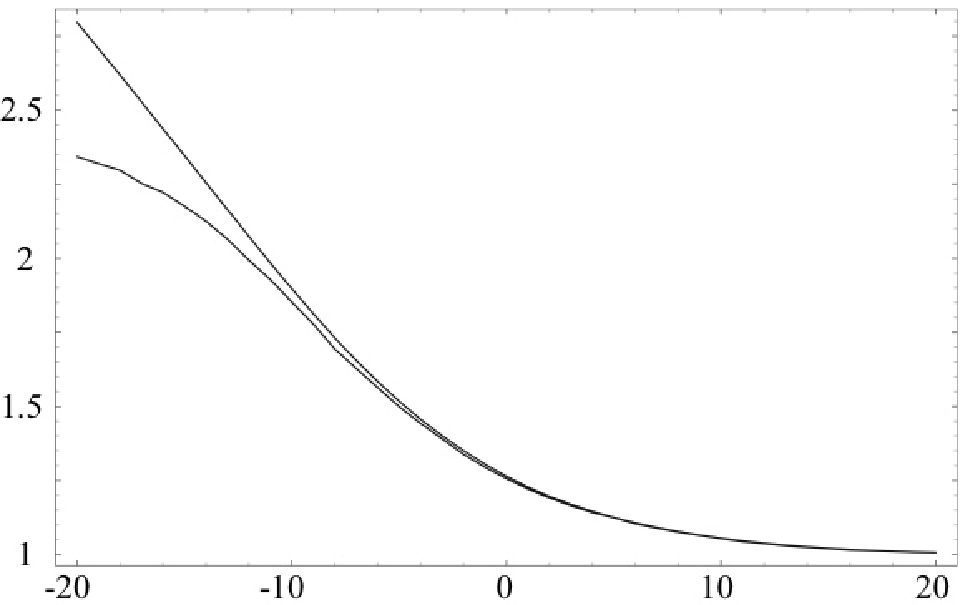}}
    \put(40,0){\makebox(0,0)[t]{\small SNR, dB}}
    \put(-1,25){\makebox(0,0)[r]{\rotatebox{90}{\small Relative
    capacity gain}}}
    \put(30,32){\makebox(0,0){{\small Space-Time}}}
    \put(15,18){\makebox(0,0){{\small Space}}}
  \end{picture}
  \caption{Rayleigh channel, $t=r=4$.}
  \label{48831fig:grel4}
\end{figure}

\begin{example}[Rayleigh $t=r=4$]
  \figref{48831fig:grel4} shows the relative capacity gain over
  $Q=PI/t$ for a four-input, four-output system. Obviously the
  additional gain over spatial water-filling is decreased compared to
  the $t=r=2$ case. In fact as $t,r \rightarrow\infty$, there is
  asymptotically no extra gain to be found by additionally
  water-filling over time as well as space. As the dimension
  increases, the eigenvalue density converges to the well-known limit
  law, holding on a per-symbol basis. Thus space-time water filling on
  Rayleigh channels is of most importance for small systems.
\end{example}

\begin{figure}[htbp]
  \centering\setlength{\unitlength}{1mm}
  \begin{picture}(80,50)
    \put(0,0){\includegraphics*[width=80mm]{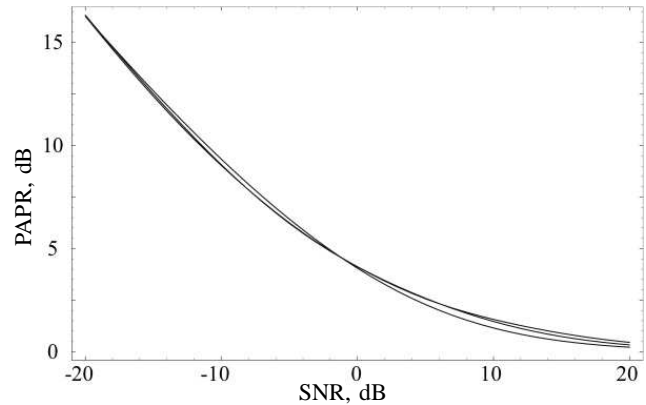}}
    \put(40,0){\makebox(0,0)[t]{\small SNR, dB}}
    \put(-1,25){\makebox(0,0)[r]{\rotatebox{90}{\small PAPR, dB}}}
  \end{picture}
  \caption{Peak-to-Average Power Ratio $t=r=1,2,4$.}
  \label{48831fig:papr}
\end{figure}

\figref{48831fig:papr} shows the peak-to-average power ratio in
decibels for $t=r=1,2,4$. Note that this is the exact value of the
PAPR. For Rayleigh channels with finite $m$, the bound of
Theorem~\ref{48831th:papr} does not apply, since $E[1/\lambda]$ does not
exist. From this figure, the peak-to-average power is relatively
insensitive to the system dimensions for the Rayleigh channel. The
particular values of PAPR are comparable with what may be experienced
in an orthogonal frequency division multiplexing system.

\begin{figure}[htbp]
  \centering\setlength{\unitlength}{1mm}
  \begin{picture}(80,50)
    \put(0,0){\includegraphics*[width=80mm]{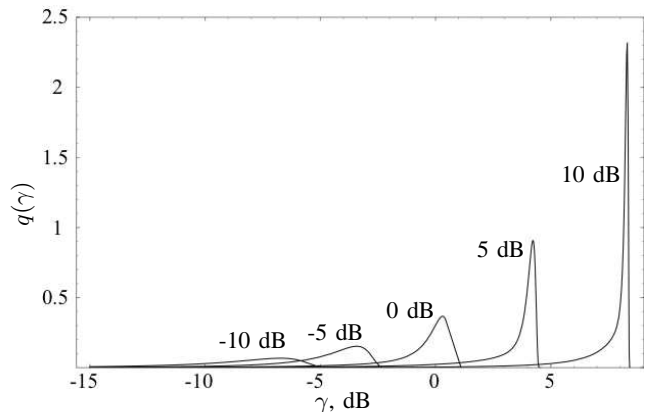}}
    \put(40,0){\makebox(0,0)[t]{\small $\gamma$, dB}}
    \put(-1,25){\makebox(0,0)[r]{\rotatebox{90}{\small $q(\gamma)$}}}
    \put(73,30){\makebox(0,0)[t]{\small 10 dB}}
    \put(61,20){\makebox(0,0)[t]{\small 5 dB}}
    \put(49,12){\makebox(0,0)[t]{\small 0 dB}}
    \put(39,9){\makebox(0,0)[t]{\small -5 dB}}
    \put(28,8){\makebox(0,0)[t]{\small -10 dB}}
  \end{picture}
  \caption{Per-eigenvector transmit power density, $t=r=2$.}
  \label{48831fig:pdist}
\end{figure}

As described earlier, the peak-to-average power ratio may be
misleading, since it is conceivable that the peak power may only be
transmitted infrequently. \figref{48831fig:pdist} shows the probability
density function of the power transmitted per-eigenvector for
$t=r=2$. At low SNR, the density is broad and has significant mass
above the target average power $P/m$. As the SNR increases, the
density converges to an impulse at $P/m$.

\section{Statistical Transmitter Side Information}\label{sec:TxStatisticalCSI}
It is tempting to think that $Q=I/\nt$ is optimal when the transmitter
has no knowledge about the channel, and assertions to this effect have
appeared in the literature.  In the complete absence of transmitter
side information however (i.e. the transmitter does not even know
$p_H$), the underlying information theoretic problem is difficult to
define. There are several possibilities, for instance $p_H$ may
selected somehow randomly from a set of possible channel
densities. Alternatively, $p_H$ could be fixed, but unknown, in the
spirit of classical parameter estimation. In the absence of a thorough
problem formulation and corresponding analysis, it is clear that
optimality of $Q=I/\nt$ is at best conjecture. For example, in the
case where $p_H$ is drawn randomly from a set of possible densities,
it may be an outage probability that is of interest. This problem is
not completely solved even when the $p_H$ are degenerate (i.e. the
non-ergodic channel of Telatar), and in that case transmission on a
subset of antennas is believed to be optimal.  We do not consider
these more difficult problems, and restrict attention to transmitter
knowledge of $p_H$.

The result \eqref{E:1.1} arises from \cite[Theorem 1]{Telatar1199} and
holds for independent, identically distributed, circularly symmetric
Gaussian channel matrix $H$, independent of transmit symbols. In
general, $Q = I_\nt/\nt$ is not optimal, and thus provides only a
lower bound to capacity. Several authors~\cite{LebFaulShafi:ICC04}
have investigated the scenario of transmitting, equal power,
independent Gaussian signals for various correlated central and
non-central random matrix channels. Other work~\cite{BocheJors1004}
have examined \emph{worst-case} mutual information in the absence of
transmitter side information, while~\cite{PaloCioffi03} has applied
game-theoretic analysis to the problem of equal power transmission,
observing that (in the absence of any better option) uniform power
allocation is not ``so bad.''

In the previous section, we considered the optimal transmit covariance
for perfect transmitter side information. We shall now relax this
constraint, so the transmitter has statistical side information only,
which is a well-posed information theoretic problem.

There are two main areas of interest. Firstly, in some scenarios, the
eigenvectors of the optimal input covariance $Q$ can be determined
a-priori (typically by inspection). Several authors have described
optimization of input covariance, by diagonalization of the transmit
covariance
\cite{JafVishGold:ICC01,JafGold:ISIT03,AlfTulLozVer:ISSSTA04}. In
other work, \cite{SimMouJSAC03} has outlined optimality conditions
for beamforming vs MIMO diversity. Recent work \cite{JorsBoch04WCOM}
has also investigated the case where input and channel covariance
matrices are jointly diagonalizable.

The more general case, is when the eigenvectors of the optimal input
covariance structure are not apparent a-priori, and may in fact be
complicated functions of $p_H$. This is the main area of interest in
this paper, and Theorem~\ref{Th:1.opto} (and the resulting iterative
optimization procedure) is our the main result. We will begin in
Section~\ref{sec:anySNR} by finding the optimal $Q$ in the
diagonalizable case, which results in an interesting comparison to
water-filling. Section~\ref{sec:MainResult} extends the result to
arbitrary $p_H$.

\subsection{Diagonalizable Covariance}\label{sec:anySNR}
Solution of Problem~\ref{Prob:1a} is in general a semidefinite
program, since the maximization is over the cone of positive
semidefinite hermitian matrices $Q \geq 0$. In certain cases however,
the problem simplifies, and we can obtain convenient conditions for
optimality from the Kuhn-Tucker conditions.  The simplest case, case
$S \sim \normal{r}{t}{0}{I\otimes I}$ was solved in
\cite{Telatar1199}. Other special cases have been solved in
\cite{JafVishGold:ICC01,JafGold:ISIT03}. Independent work finding
similar results to those described below has appeared in
\cite{TulLozVerdu0705}.

Suppose it can be determined that the optimal $Q$ has the form
\begin{align}
  Q &= U\hat{Q}\herm{U} \label{eq:diagQ}\\
  \hat{Q} &= \diag\left(q_1, q_2,\dots,q_\nt\right) \label{eq:diagQ1a}
\end{align}
for some fixed $U$.  For such channels, the optimization problem
reduces to finding the best allocation of power to each column of $U$.

One important example is
$H[\symboltime]\sim\normal{m}{m}{0}{\sr\otimes\st}$, i.e. the
Kronecker correlated Rayleigh channel with no line-of-sight
components. In that case, is is known that $U$ diagonalizes $\st$ and
optimal transmission is independent on each eigenvector of $\st$.

In such cases, the condition $Q>0 \implies \hat{Q}>0$ allows the
application of the Kuhn-Tucker conditions for maximization of a convex
function over the space of probability vectors~\cite[p.
87]{Gallager68} to yield the following lemma.
\begin{lemma}\label{D:lagrange} Consider
  the channel~\eqref{E:1.the_channel} with
  $H[\symboltime]\sim\normal{m}{m}{0}{\sr\otimes\st}$.  The optimal
  covariance $Q$ has the form~\eqref{eq:diagQ} and satisfies the
  Kuhn-Tucker conditions~\cite[p. 87]{Gallager68}
  \begin{align}
    \frac{\partial \Psi(Q)}{\partial q_i} &= \mu \quad q_i > 0 \label{eq:KT1}
    \\
    \frac{\partial \Psi(Q)}{\partial q_i} &\leq \mu \quad q_i = 0  \label{eq:KT2}
  \end{align}
%\begin{align}
%\frac{\partial f(\nu)}{\partial \nu_{ij}} &= 2\mu \nu_{ij}, \quad i\neq j, \mu>0\\
%&=2\mu \nu_{ii}, \quad \nu_{ii} > 0, \mu>0\\
%&<0\quad \nu_{ii} = 0
%\end{align}
  where $\mu$ is a constant independent of $q_i$, and $q_i$ are given by~\eqref{eq:diagQ1a}.
\end{lemma}
Thus the necessary and sufficient conditions for optimality have a
particularly simple form. Differentiating $\Psi(Q) =
\expect[H]{\log\det\left(I+HQ\herm{H}\right)}$ leads to the following
theorem, proved in~\cite{HanGra:04IT-unpub}.

\begin{theorem}[Optimal Covariance]\label{Th:cap_distribute}
  Consider the ergodic channel~(\ref{E:1.the_channel}) with $p_H$ such
  that the optimal input covariance is known to be of the form
  \eqref{eq:diagQ}-\eqref{eq:diagQ1a} for some fixed unitary matrix
  $U$.  A necessary and sufficient condition for the optimality of the
  diagonal $\hat{Q}$ in~\eqref{eq:diagQ1a} is
  \begin{align}
      \expect[S]{\left(\left(I+S\hat{Q}\right)^{-1}S\right)_{kk}}
      = \mu && q_k > 0 \label{E:thm_opt_cov} \\
      \expect[S]{\left(\left(I+S\hat{Q}\right)^{-1}S\right)_{kk}} < \mu &&
      q_k=0 \label{E:thm_opt_cov0}
  \end{align}
  for $k=1,2,\dots, \nt$ and some constant $\mu$. The expectation is
  with respect to the random matrix $S = \snr\,\herm{U}\herm{H}HU$.
%, where
%  $H\sim\normal{\nr}{\nt}{0}{\sr\otimes\st}$ and $U$ diagonalizes $T$.
  The notation $\left(A\right)_{ij}$ denotes element $ij$ of $A$.
\end{theorem}

In the case $Q>0$, the condition~\eqref{E:thm_opt_cov} may be
re-written as a fixed-point equation
\begin{equation}\label{E:opt_cov2}
  \hat{Q} = \nu\,\expect[S]{\left(\hat{Q}^{-1}+S\right)^{-1} S},
\end{equation}
which suggests the following iterative procedure for numerically
finding the optimal $\hat{Q}$.
Starting from an initial diagonal $\hat{Q}^{(0)}>0$, compute
\begin{equation}
  q_k^{(i+1)} =
  \nu^{(i)}\left[\expect[S]{\left((\hat{Q}^{(i)})^{-1} + S
        \right)^{-1}S}\right]_{kk},  
\end{equation} 
selecting $\nu^{(i)}$ at each step to keep $\tr \left(\hat{Q}^{(i)}\right)=\snr$.
Although there is no known closed form solution for
$\expect[S]{\left(\hat{Q}^{-1}+S\right)^{-1}S}$, it may be accurately
estimated using monte-carlo integration. Note that the numerical
procedure may be applied to each entry $q_k = Q_{kk}$ separately for a
given $\hat{Q}^{(i)}$. Numerically, each fixed point iteration is
performed once and the $\nt$ non-zero diagonal entries of $\hat{Q}$
are updated.

It is interesting to compare the conditions~\eqref{E:thm_opt_cov},
\eqref{E:thm_opt_cov0} with the solution of Problem~\ref{Prob:1}, for
perfect transmitter side information. Suppose $H[i]=H$ is known at the
transmitter with $H\herm{H}=US\herm{U}$ being the eigenvalues
decomposition of $H\herm{H}$. The Kuhn-Tucker condition for optimality
of the input covariance $Q=\herm{U}\hat{Q}\herm{U}$ can be written in
the following form,
\begin{align}
  \left(\left(I+S\hat{Q}\right)^{-1}S\right)_{kk} \label{E:opt_cov_qkp}
  = \mu && q_k > 0 \\
  \left(\left(I+S\hat{Q}\right)^{-1}S\right)_{kk} < \mu &&
  q_k=0. \label{E:opt_cov_qkz}
\end{align}
with $\hat{Q}$ satisfying \eqref{eq:diagQ1a}. Solution of these
equations is straightforward and leads easily
to~\eqref{48831eq:capacityC} and \eqref{48831eq:capacityP}.

Comparing \eqref{E:thm_opt_cov} with \eqref{E:opt_cov_qkp} it can be
seen that the \emph{only} difference is the presence of the
expectation in~\eqref{E:thm_opt_cov}.  Similarly for
\eqref{E:thm_opt_cov0} and \eqref{E:opt_cov_qkz}. This is no real
surprise, and is due to the interchangability of differentiation and
expectation. The result of Theorem~\ref{Th:cap_distribute} is a direct
generalization of the classical water-filling result for parallel
channels~\cite{CairSham091999}, where the transmitter has statistical
side information, and the channel can be diagonalized
a-priori. In the latter case however, there is no water-filling
interpretation~\cite{GuoShaVer:IT05}.

For the deterministic case, it is clear that increasing $\snr$ can only increase the power allocated to any particular eigenvector (water-level raises). The same thing happens in the ergodic case, as demonstrated by the following theorem, proved in the Appendix.
\begin{theorem}\label{th:monotonic}
Let $\hat{Q}=\diag(q_1,\dots,q_\nt)$ be the eigenvalues of the optimal covariance matrix for a channel with signal-to-noise ratio $\snr$, satisfying the conditions of Theorem \ref{Th:cap_distribute}. Then
\begin{equation*}
\frac{\partial q_k}{\partial\snr} \geq 0, \quad k=1,2,\dots,\nt.
\end{equation*}
\end{theorem}
Thus a signal-to-noise ratio increase (decrease) can only increase (decrease) the power allocated to each eigenvector of the optimal covariance matrix.

Theorem \ref{Th:cap_distribute} takes care of zero-mean Rayleigh
fading channels with separable correlation structure.  In the case of
Ricean fading with non-zero mean, one approach is to use the following
approximation by a central distribution.
\begin{lemma}[Wishart Approximation~\cite{GuptaNagar99}]\label{lem:appwish}
  Suppose $H\sim\normal{\nr}{\nt}{M}{I\otimes\st}$. Then $S = H Q
  \herm{H}$ may be approximated by a \emph{central} Wishart
  matrix~\cite[p. 125]{GuptaNagar99}
  \begin{align}
    S &\sim \wishart{\nt}{0}{\Sigma} \label{E:wishart_approx}
    \\ \Sigma &= \st^{1/2}Q\st^{1/2} + \frac{1}{\nt}\herm{M}M \label{E:wishart_approx_recipe}
  \end{align} 
\end{lemma}
This approximation motivates application of Theorem
\ref{Th:cap_distribute} to the Ricean case with
$H[\symboltime]\sim\normal{\nr}{\nt}{M}{I\otimes\st}$ . The relation
between correlation and line-of-sight (non-zero mean) has been
heuristically established in MIMO channel measurement
literature~\cite{KyrCoxValWol03,Yu0501,Yu1102}. The accuracy of this
approximation is investigated numerically below.

\begin{figure}
  \centering\setlength{\unitlength}{1mm}
  \begin{picture}(80,80)
    \put(0,0){\includegraphics*[width=80mm]{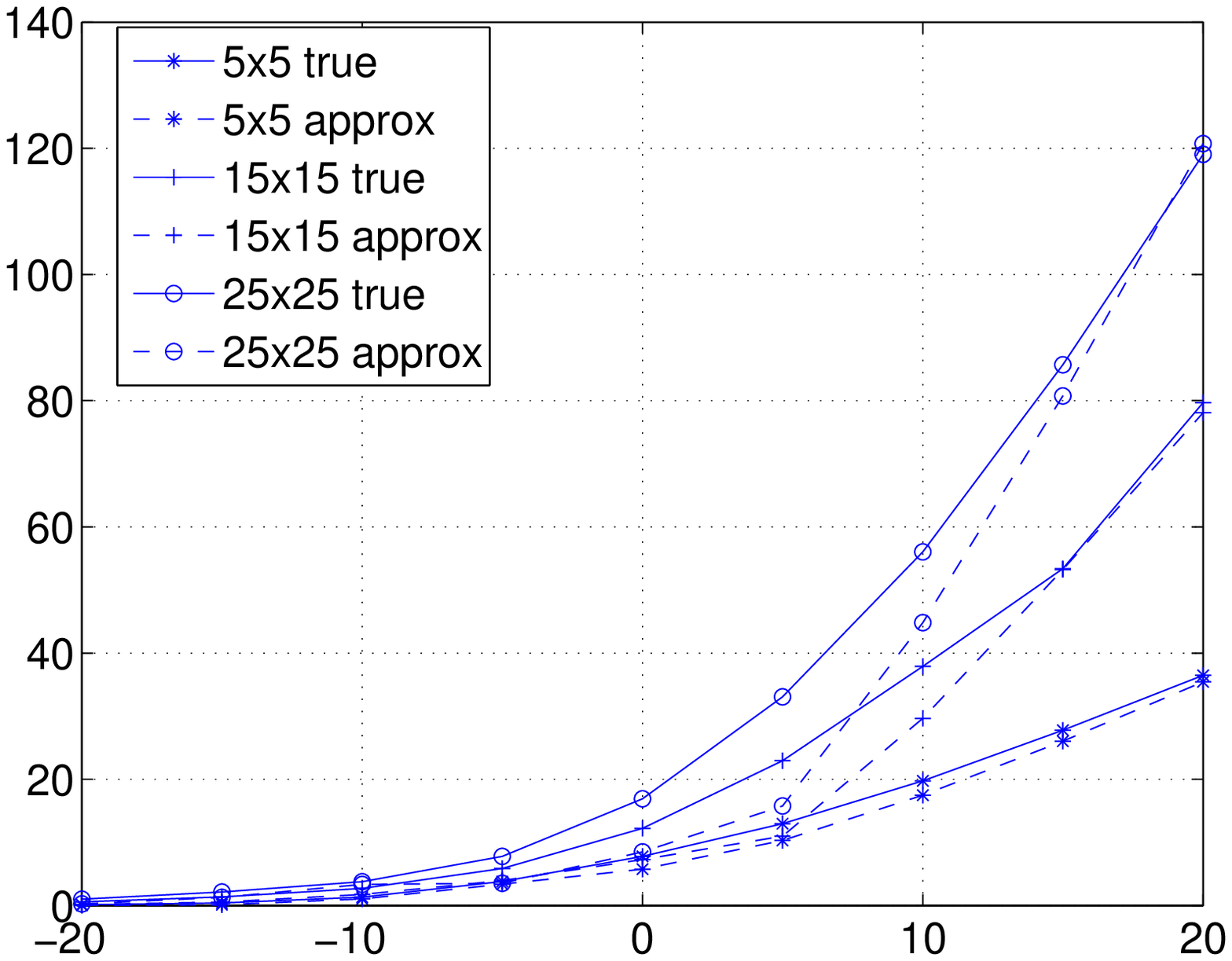}}
    \put(40,-1){\makebox(0,0)[t]{\small SNR (dB)}}
    \put(0,30){\makebox(0,0)[r]{\rotatebox{90}{\small mutual information}}}
  \end{picture}
  \caption{Mutual Information $I$ with central Wishart approximation, for non-central channel. Solid lines give $I(Q)$ (capacity) for optimal input covariance, while dashed lines give $I(Q_a)$ where $Q_a$ is optimal according to a central Wishart approximation. Closest results are given at high- and low-SNR and small numbers of elements.}
  \label{wishart:approx:fig}
\end{figure}

In figure~\ref{wishart:approx:fig} we have plotted the capacity and the mutual information for a channel with rank-one mean $M=\diag\{\nt,0,\ldots,0\}$ and non-diagonal transmit covariance 
\begin{equation}\label{E:Tdef:ones}
T=\begin{bmatrix}1 &\tau &\tau &\cdots\\\
\tau &1 &\tau &\cdots\\
\vdots & &\ddots \end{bmatrix}
=\tau \ones +  \diag\{\tau-1\}
\end{equation}
where $\ones$ is a matrix of all ones.

 The plot compares the capacity (optimal input covariance, with \emph{true} probability law) with the mutual information (input covariance given by central Wishart \emph{approximation}) for various SNR and numbers of transmit and receive elements. Each plot has assumed $\nt=\nr$. We note that the  approximated covariance matrix is a linear combination of the transmit-end covariance $T$ and the mean, and thus approximated input covariance is a dominated by beamforming on $M$ at low SNR, and $T$ at higher SNR. 

%Consider a MIMO system with $\nt=\nr=n$ transmit and receive elements in Rician fading.
%  Let $$H = \sqrt{\frac{\kappa}{\kappa+1}}M +
%  \sqrt{\frac{1}{\kappa+1}}\sr^{1/2}X.$$  From
%  \eqref{E:wishart_approx_recipe}, create an approximate correlated
%  Rayleigh model, for an approximation to capacity:
%  $$H \approx \sqrt{\frac{1}{\kappa+1}}\left(\sr^{1/2} +
%    \sqrt{\frac{\kappa}{n^2}}I\right) X$$
%The optimal input covariance may be found from Theorem \ref{}, with
%appropriate substitutions.

Beamforming, i.e. rank-one transmission with
$\hat{Q}=\diag(1,0,0,\dots,0)$ is a particularly simple strategy,
which is optimal at low SNR (see Section \ref{sec:asymptotics}). It is
interesting to consider the conditions under which beamforming is
optimal.
\begin{theorem}\label{th:optbeam}
  Consider an ergodic channel (\ref{E:1.the_channel}) with
  $H\sim\normal{\nt}{\nr}{0}{\sr \otimes \st}$, where without loss of
  generality $\sr=\diag(\sri_1,\dots,\sri_\nr)$ and
  $\st=\diag(\sti_1,\dots,\sti_\nt)$ with $\tr(\st)=\nt$ and
  $\tr(\sr)=\nr$. Beamforming is optimal if and only if
\begin{equation}\label{eq:optbeam}
   \expect{\frac{ \herm{u}\sr u + \snr\sti_k
       \herm{u}\sr^2u}{1+\snr\sti_1 \herm{u}\sr u}}  
  \geq \nr \frac{\sti_k}{\sti_1} \quad \text{for any}\ k\geq 2,
\end{equation}
where the expectation is with respect to a length $\nr$ Gaussian
vector with i.i.d. unit variance entries,
$u\sim\normal{\nr}{1}{0}{I}$.

% \begin{equation*}
%   p_W(W) = \frac{1}{2^{2\nr} \tilde{\Gamma}_2(\nr)} \det(\sr)^{-2}
%   \det(W)^{\nr-2} \mbox{}_0F_0\left(\sr^{-1},-W\right). 
% \end{equation*}

\end{theorem}

The left hand side of \eqref{eq:optbeam} is monotonically decreasing with signal-to-noise ratio.

%The density $p_W(W)$ may be compactly represented as a function of the
%eigenvalues of $W$, however \eqref{eq:optbeam} cannot be expressed
%easily in terms of those eigenvalues. Nevertheless, the expectation
%with respect to $p_W$ may be performed numerically using the
%determinant expression \cite{GroRich89} for $\mbox{}_0F_0$.

For zero-mean Rayleigh channels, the condition
\eqref{eq:optbeam} can be found in closed form \cite{SimMou03}. In the
appendix we give an alternate proof to that given by
\cite{SimMou03}. Our proof is simplified via use of Theorem
\ref{th:optbeam}.
\begin{theorem}[Simon and Moustakas \cite{SimMou03}]\label{th:optbeam2}
  Consider an ergodic channel (\ref{E:1.the_channel}) with
  $H\sim\normal{\nt}{\nr}{0}{\sr \otimes \st}$, where without loss of
  generality $\sr=\diag(\sri_1,\dots,\sri_\nr)$ and
  $\st=\diag(\sti_1,\dots,\sti_\nr)$ with $\tr(\st)=\nt$ and
  $\tr(\sr)=\nr$. Beamforming is optimal if and only if
  \begin{equation}
    \sum_{i,j=1}^\nr
    \frac{\sri_i(1+\snr\sti_2\sri_i)\sri_j^{\nr-1}}{\prod_{k\neq
      j}(\sri_j-\sri_k)} {\zeta_{ij}} >
  \nr \snr \sti_2 
  \end{equation}
where
\begin{align*}
  \zeta_{ij} &=
  \begin{cases}\displaystyle
    \frac{f(\snr\sti_1\sri_i)-f(\snr\sti_1\sri_j)}{\sri_i-\sri_j}& i\neq j \\ 
    \displaystyle \frac{1}{\sri_i} \left(1 - \frac{f(\snr\sti_1\sri_i)}{\snr\sti_1\sri_i}\right)&  i=j
  \end{cases} \\
  f(x) &= e^{1/x} \Gamma(0,1/x).
\end{align*}
\end{theorem}
In the above theorem, note that $\zeta_{ii}$ is just the limit of
$\zeta_{ij}$ as $\sri_i\to\sri_j$.
Theorem \ref{th:optbeam} is a generalization of \cite{JafGol01ISIT}
(which was for the MISO case), and the MISO result is recovered easily
from \eqref{eq:optbeam} via $\nr=1$.

\begin{figure}%[htbp]
  \centering\setlength{\unitlength}{0.88mm}
  \begin{picture}(90,90)
    \put(0,0){\includegraphics*[width=70mm]{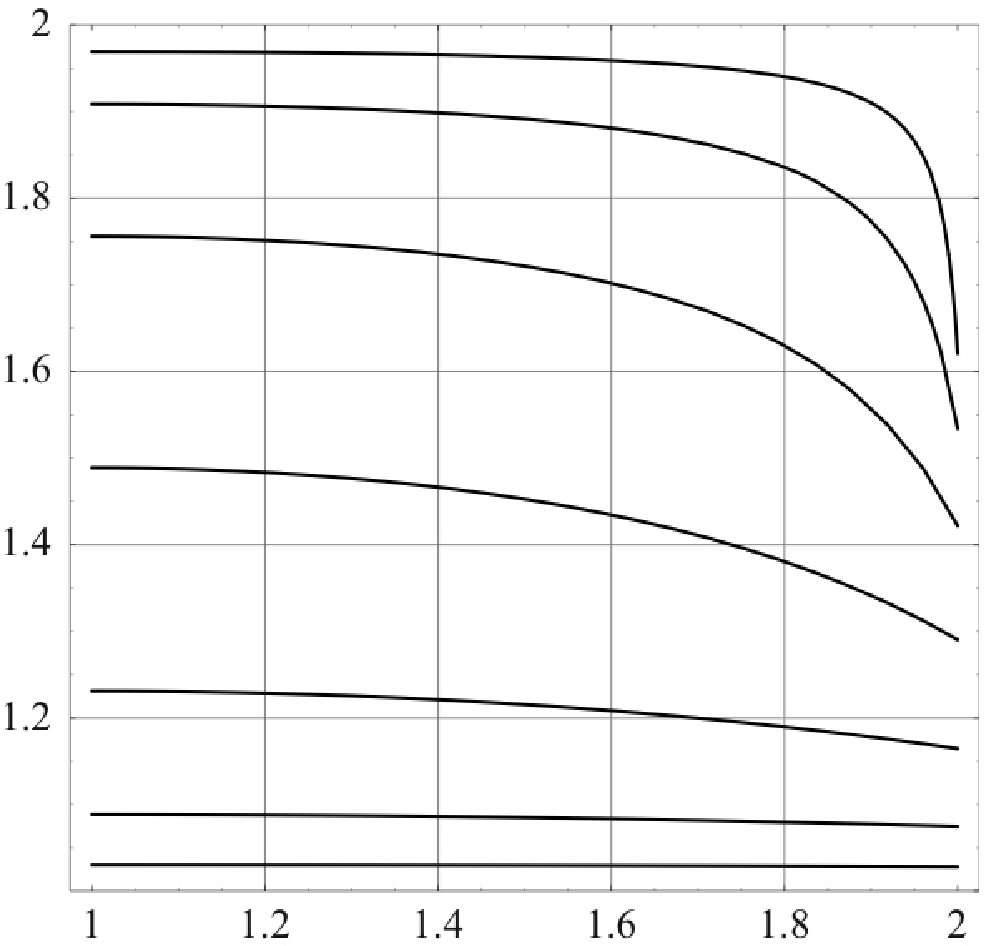}}
    \put(40,-1){\makebox(0,0)[t]{\small $\rho$}}
    \put(-1,40){\makebox(0,0)[r]{\rotatebox{90}{\small $\tau$}}}
    \put(15,9){\makebox(0,0)[t]{\tiny -15dB}}
    \put(20,13){\makebox(0,0)[t]{\tiny -10dB}}
    \put(25,23){\makebox(0,0)[t]{\tiny -5dB}}
    \put(35,36){\makebox(0,0)[t]{\tiny 0dB}}
    \put(50,53){\makebox(0,0)[t]{\tiny 5dB}}
    \put(63,62){\makebox(0,0)[t]{\tiny 10dB}}
    \put(69,68){\makebox(0,0)[t]{\tiny 15dB}}
  \end{picture}
  \caption{Optimality of beamforming. Beamforming is optimal for a given SNR for all points $(\tau,\rho)$ above the line corresponding to that SNR value. The plot is symmetric for $0\leq\rho<1$ and $0\leq\tau<1$}
  \label{optbeam:fig}
\end{figure}

%\begin{example}
Figure \ref{optbeam:fig} shows the beamforming optimality condition of Theorem~\ref{th:optbeam2} for a set of SNR levels $\snr$ and a $2\times2$ channel, with $H\sim\normal{2}{2}{0}{\sr\otimes \st}$ where $\sr=\diag\{\sri, 2-\sri\}$ and $\st=\diag\{\sti, 2-\sti\}$,  $1\leq \sri,\sti \leq 2$. The plot is symmetric around the point $\sri=\sti=1$ (and thus, only the top-left quadrant of the full $0\leq\sti,\sri\leq2$ plot is shown).

The lines provide the transition point from regions where beamforming is optimal (above each line) to regions where beamforming is not optimal. The plot shows the region for $1\leq\sti,\sri\leq 2$. For $\sti=1$, $\st=I$ and for $\sti\to 2$, $\st$ becomes singular, similarly for $\sr$: so that the top right-hand corner of the plot has highly correlated $H$, whilst the bottom left-hand corner has iid $H$.

It can be seen that for low SNR, $\gamma=-15$dB, beamforming is almost always optimal with the transition occurring for $\sti \approx 1.03$. Note also, that the eigenvalues of $\sr$ have little effect on the optimality of beamforming at low SNR. As SNR increases, the region for admissible covariance matrices for optimal beamforming reduces: we require more covariance matrices with larger eigenvalue separation. The optimality of beamforming is clearly dependent upon the eigenvalues of $\st$. At higher SNR, the optimality of beamforming is also dependent on $\sr$ (as can be seen by the $\gamma\geq0dB$ curves. The reason for this is that the low rank of $\sr$ results in an effective power loss at the receiver.
%\end{example}

\subsection{The General Case}\label{sec:MainResult}
We now wish to solve Problem \ref{Prob:1a}, \emph{without the a-priori
  requirement of diagonal input covariance}. In this case, we need to
maximize $\psifunc{Q}{S}$ over all positive definite $Q$.  In
particular we do not wish to restrict ourselves to particular matrix
densities such as the zero-mean Kronecker Gaussian model. 

Whilst of interest in its own right, this problem arises when the
input covariance structure cannot be solved by inspection. Specific
examples include the non-central Gaussian random matrix channel, where
the channel covariance and mean are not jointly diagonalizable, and
for several random matrix channels which do not have simple
(Kronecker) factorizations~\cite{OzcCziBon05,OzcHer03EL}.

%% \begin{equation}
%% H = R\odot X_{iid}
%% \end{equation}
%% with $H,R,X\in\complex^{\nr\times\nt}$. This channel has recently been
%% examined in

To accommodate the positive definite constraint on $Q$, we apply the
Cholesky factorization, so the constraint becomes implicit in the
final solution. By adopting this approach we force the optimization to
only consider the minimum number of independent variables required for
solution, $t(t+1)/2$ rather than $t^2$. 

Any non-negative matrix $A$ may be written as~\cite{Muirhead82}
\begin{equation}\label{E:triangle-ident}
  A = \herm{\trimat} \trimat
\end{equation}
for upper triangular matrix $\trimat$, with the diagonal elements $\trimati_{ii}$
real and non-negative. Similarly, for a given upper triangular matrix
$\trimat$, the product $\herm{\trimat}\trimat$ is positive definite. The following
useful properties~\cite{GuptaNagar99} arise from
\eqref{E:triangle-ident}, $\trace(A) = \trace(\herm{\trimat}\trimat) = \sum_{i\leq
  j}\trimati_{ij}^2$ and $\det(A)=\prod_{i}\trimati_{ii}^2$.

Using \eqref{E:triangle-ident}, transform Problem~\ref{Prob:1a} to
\begin{problem}[Equivalent to Problem~\ref{Prob:1a}]\label{Prob:2}
  \begin{align*}
    \max_{\trimat}&\, \Psi(\herm{\trimat}\trimat, p_H) \\
    \intertext{subject to}
    \sum_{i\leq j}\trimati_{ij}^2 &= 1 \\
     \trimati_{ii} &\geq  0,\quad  \forall i
  \end{align*}
\end{problem}
The maximum $\Psi^o$ for optimal $\trimati^o$, is not improved by choosing a
trace less than unity, hence equality of the first constraint.

Problem~\ref{Prob:2} admits a quadratic optimization approach, using
Lagrange multipliers~\cite{BoydVanden04}. The optimization in
Problem~\ref{Prob:2} occurs on the (upper triangular) matrix $T$ which
has \emph{exactly} $t(t+1)/2$ independent (complex) variables.  This corresponds
to the number of independent variables for the optimization over $Q$
in Problem~\ref{Prob:1}, since $Q=U\hat{Q}\herm{U}$ has $t$ independent
variables in the diagonal matrix $\hat{Q}$ and $t(t-1)/2$ independent
variables in the unitary matrix $U$.

In order to solve Problem~\ref{Prob:2}, we produce a modified cost function $J(\nu,\mu,\phi)$ where $\nu=\vec{\trimat}$, $\mu$ and $\phi$ are vectors of Lagrange multipliers corresponding to equality and inequality constraints. For this we use the following:
\begin{lemma}[Application Kuhn-Tucker Theorem~\cite{KuhnTucker51}]\label{Lem:1}
  Given a convex $\cap$ function $f(\nu)$ of a vector $\nu$, where $\nu$ is  constrained by:
\begin{equation*}
\sum_{i<j}\nu_{ij}^2 = 1 \quad
\nu_{ii} \geq 0
\end{equation*}
then
%\begin{align}
%\frac{\partial f(\nu)}{\partial \nu_{ij}} &= 2\mu \nu_{ij}  
%    \quad \nu_{ij}>0,  \label{E:kteq}\\
%   &<0 \quad i=j, \nu_{ii}=0 \label{E:ktneq1}
%\end{align}
\begin{align}
\frac{\partial f(\nu)}{\partial \nu_{ij}} &= 2\mu \nu_{ij}, \quad i\neq j, \mu>0\\
&=2\mu \nu_{ii}, \quad \nu_{ii} > 0, \mu>0\\
&<0\quad \nu_{ii} = 0
\end{align}
defines a maximum point for the function $f(\nu)$.
\end{lemma}

Lemma~\ref{Lem:1} provides the necessary conditions for a vector $\nu=\mathrm{vec}(T)$ to give a capacity achieving input covariance. We now present the main result of the paper: a general condition for the capacity achieving input covariance.

\begin{theorem}[Optimal Transmit Covariance]\label{Th:1.opto}
  Given a MIMO channel~(\ref{E:1.the_channel}) with the channel chosen
  ergodic according to a probability distribution $p_H$, then the
  capacity achieving input is Gaussian with covariance $Q=\herm{\trimat}\trimat$
  where $\trimat$ is upper triangular, and the element $\trimati_{ij}$ satisfies:
  \begin{align}
    \expect[S]{ \trace\left[(I+S\herm{\trimat}\trimat)^{-1} SE^{(ij)} \right]} &=
    \begin{cases} 
    2\mu\trimati_{ij}  &\quad i\neq j, \mu>0 \\
    2\mu\trimati_{ii} &\quad \trimati_{ii}>0,\mu>0
    \end{cases}\label{E:thm8:fixedpt}\\
    \expect[S]{ \trace\left[(I+S\herm{\trimat}\trimat)^{-1} SE^{(ii)} \right]} &< 0 \quad \trimati_{ii} =0
  \end{align}
  where the expectation is with respect to $S=\herm{H}H$, the constant
  $\mu$ is chosen to satisfy the power constraint and
  \begin{align*}
    E^{(ij)} &= \frac{\partial \herm{\trimat}\trimat}{\partial \trimati_{ij}} \\
    \left(E^{(ij)}\right)_{mn} &= \trimati_{in}\delta_{mj} + \trimati_{im}\delta_{nj}.
  \end{align*}
with $\delta_{ij}=1$ when $i=j$ and zero otherwise.
\end{theorem}

The capacity of the channel is then given by application of $\trimat$ in
$\psifunc{\herm{\trimat}\trimat}{S}$:
\begin{equation*}
C = \expect{ \log\det\left(I_r+S\herm{\trimat}\trimat\right)}
\end{equation*}
Given the result of Theorem~\ref{Th:1.opto}, we wish to numerically
evaluate the optimal covariance, and hence capacity for an arbitrary
multiple-input, multiple-output channel.  Fortunately, the form of
\eqref{E:thm8:fixedpt} also lends itself to a fixed-point algorithm.

If we define the matrix 
\begin{equation}\label{E:def-M}
M = \expect{ (I+S\herm{\trimat}\trimat)^{-1}S }
\end{equation}
then 
\begin{equation}
\trace(ME^{(ij)}) = \sum_k (m_{kj}+m_{jk})\trimati_{ik}=\left[\trimat(M+\herm{M})\right]_{ij}
\end{equation}
The matrix $M$ may be interpreted as a differential operator, on the
function $\psifunc{\herm{\trimat}\trimat}{S}$, evaluated at a particular value of $T$.
This provides a direct fixed-point equation of projected gradient
type~\cite{Rosen61}:
\begin{equation}
\nu^{(k+1)} = -\frac{1}{\mu}\nu^{(k)}\cdot \nabla
\expect[S]{\Psi\left(\nu^{(k)}\right)} 
\end{equation}

Writing this out completely gives the following algorithm
\begin{algorithm}[Iterative Power Allocation]\label{Alg:1}\hfill\ \\
\begin{enumerate}
\item Update using \eqref{E:def-M}
\begin{equation}
\trimat^{(k+1)} \rightarrow   \trimat^{(k)} %\expect[S]{\left(I+S\herm{\trimat^{(k)}} \trimat^{(k)}\right)^{-1} S }
\left(M+\herm{M}\right)
\label{E:fixed-pt}
\end{equation}

\item Scale
\begin{equation}
\left[\trimat^{(k+1)}\right]_{ij} \rightarrow \begin{cases}
\frac{1}{\mu} \left[\trimat^{(k+1)}\right]_{ij} &i\leq j \\
0 &\text{otherwise}
\end{cases}
\end{equation}
with $\mu$ constant for all $i,j$ and chosen so that $\trace\left(\herm{\trimat}\trimat\right)=1$.

\item Repeat
\end{enumerate}
%for $i\leq j$.
\end{algorithm}

We denote $\trimat^{(k)}$ as the triangular matrix at iteration $k$. This
algorithm may be initiated with any (upper triangular) $\trimat$ satisfying
$\trace(\herm{\trimat}\trimat)=1$. The expectation \eqref{E:fixed-pt} is typically
intractable and may be evaluated using monte-carlo integration.

\begin{theorem}\label{Prop:1} Algorithm~\ref{Alg:1} converges to
  the optimal covariance $Q^o=\herm{\trimat}\trimat$. 
\end{theorem}

We note that the stability of the algorithm is directly affected by the stability of the expectation in \eqref{E:def-M}. In particular, at high-SNR, the off-diagonal entries of $\trimat$ will approach zero (since $Q=\alpha I$ is optimal). In this case, the elements of $\trimat$ may fluctuate as small movements over the Haar manifold (small changes in eigenvectors) result in large changes in the entries of $\trimat$.

\begin{figure}[tbp]
\setlength{\unitlength}{0.7mm}
\begin{center}
\begin{picture}(105,80)
  \put(0,5){\includegraphics[width=80mm]{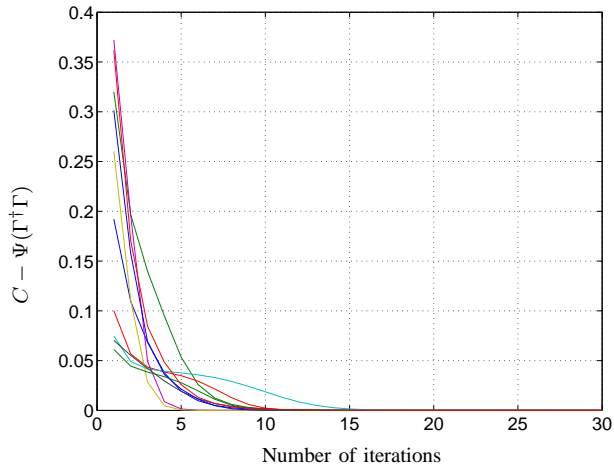}}
  \put(47,0){\footnotesize Number of iterations}
  \put(-1,30){\rotatebox{90}{\footnotesize $C-\Psi(\herm{\trimat}\trimat)$}}
\end{picture}
\end{center}
  \caption{Convergence of Algorithm~\ref{Alg:1} with $4\times4$ matrix. $S=US_o\herm{U}$ \eqref{E:So_def}. $C=1.1394$ nat/s}\label{F:converge}
\end{figure}

%\subsection{Deterministic Channel: Statistical Water-filling Algorithm Convergence}
In Figure~\ref{F:converge} we show an example of the convergence of
the algorithm for several deterministic channel matrices. Each curve
shows the difference between the mutual information for $Q=\herm{\trimat}\trimat$ vs the channel capacity $C$ for the $k^{th}$ iteration.

The example channel matrices were chosen to have common eigenvalues,
but randomly chosen eigenvectors (thus each instance has the same
capacity, but different optimal input covariance), with    
\begin{equation}\label{E:So_def}
S=US_o\herm{U}, S_o = \left(\begin{smallmatrix}2 &0\\ 0
    &1\end{smallmatrix}\right)
\end{equation}

\begin{figure}[tp]
\setlength{\unitlength}{0.7mm}
\begin{center}
\subfigure[$\nt=\nr=5$\label{F:conv:F1}]{
\begin{picture}(105,80)
  \put(0,0){\includegraphics[width=60mm]{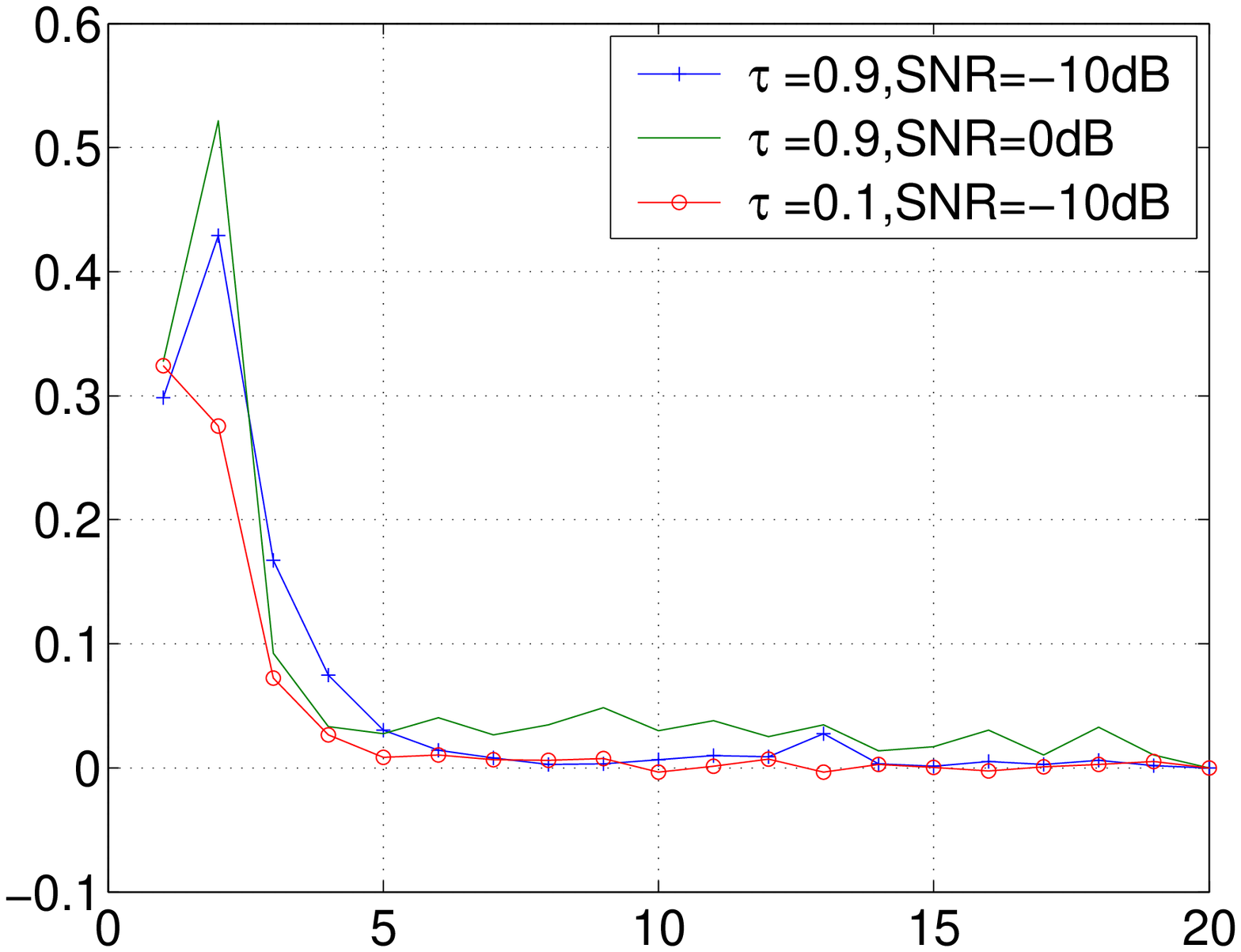}}
  \put(27,1){\footnotesize  iterations}
  \put(0,30){\rotatebox{90}{\footnotesize $C-\Psi(\herm{\trimat}\trimat)$}}
\end{picture}
}  
\subfigure[$\nt=\nr=15$\label{F:conv:F2}]{
\begin{picture}(105,80)
  \put(0,0){\includegraphics[width=60mm]{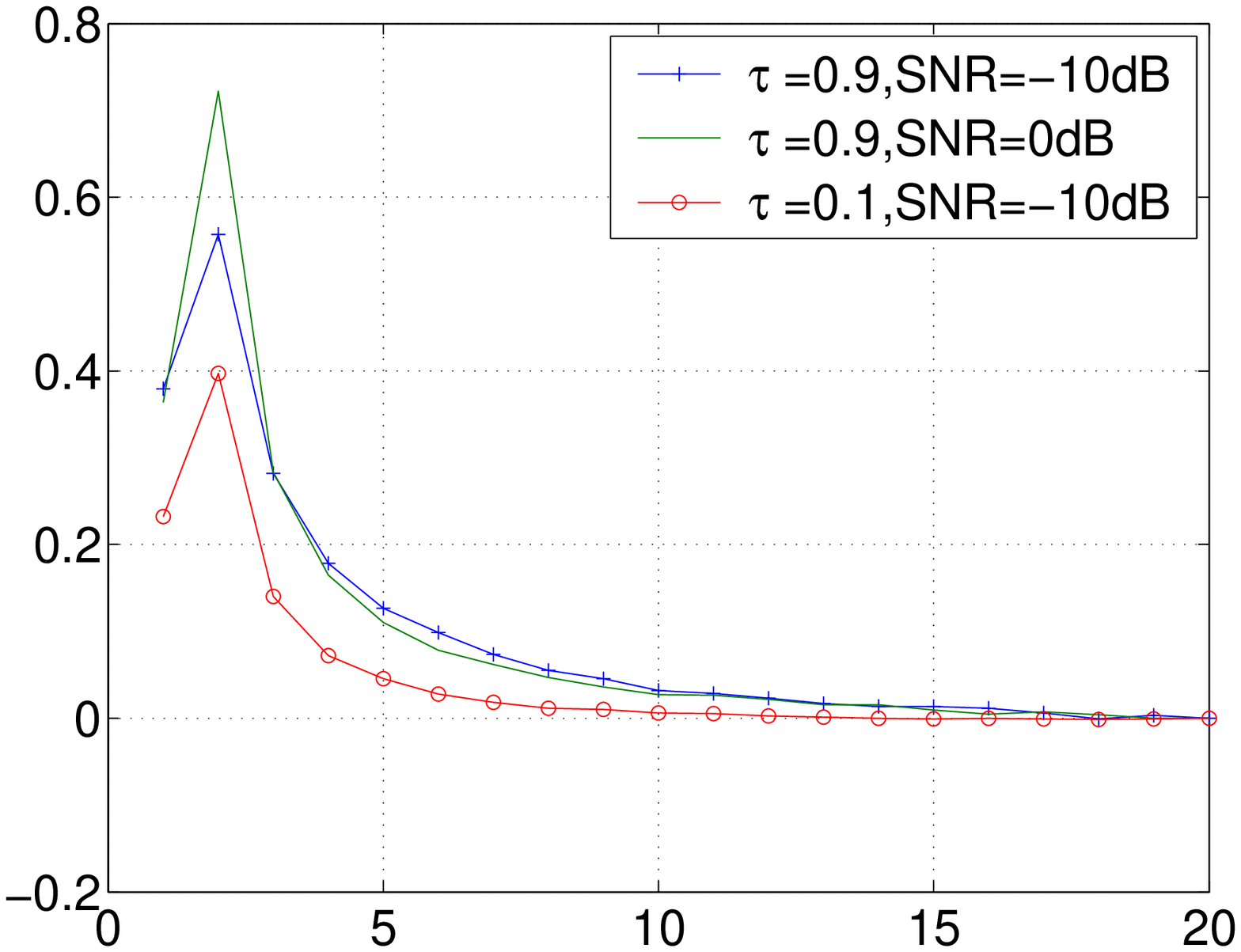}}
  \put(27,1){\footnotesize  iterations}
  \put(0,30){\rotatebox{90}{\footnotesize $C-\Psi(\herm{\trimat}\trimat)$}}
\end{picture}
}
\subfigure[$\nt=\nr=25$\label{F:conv:F3}]{
\begin{picture}(105,80)
  \put(0,0){\includegraphics[width=60mm]{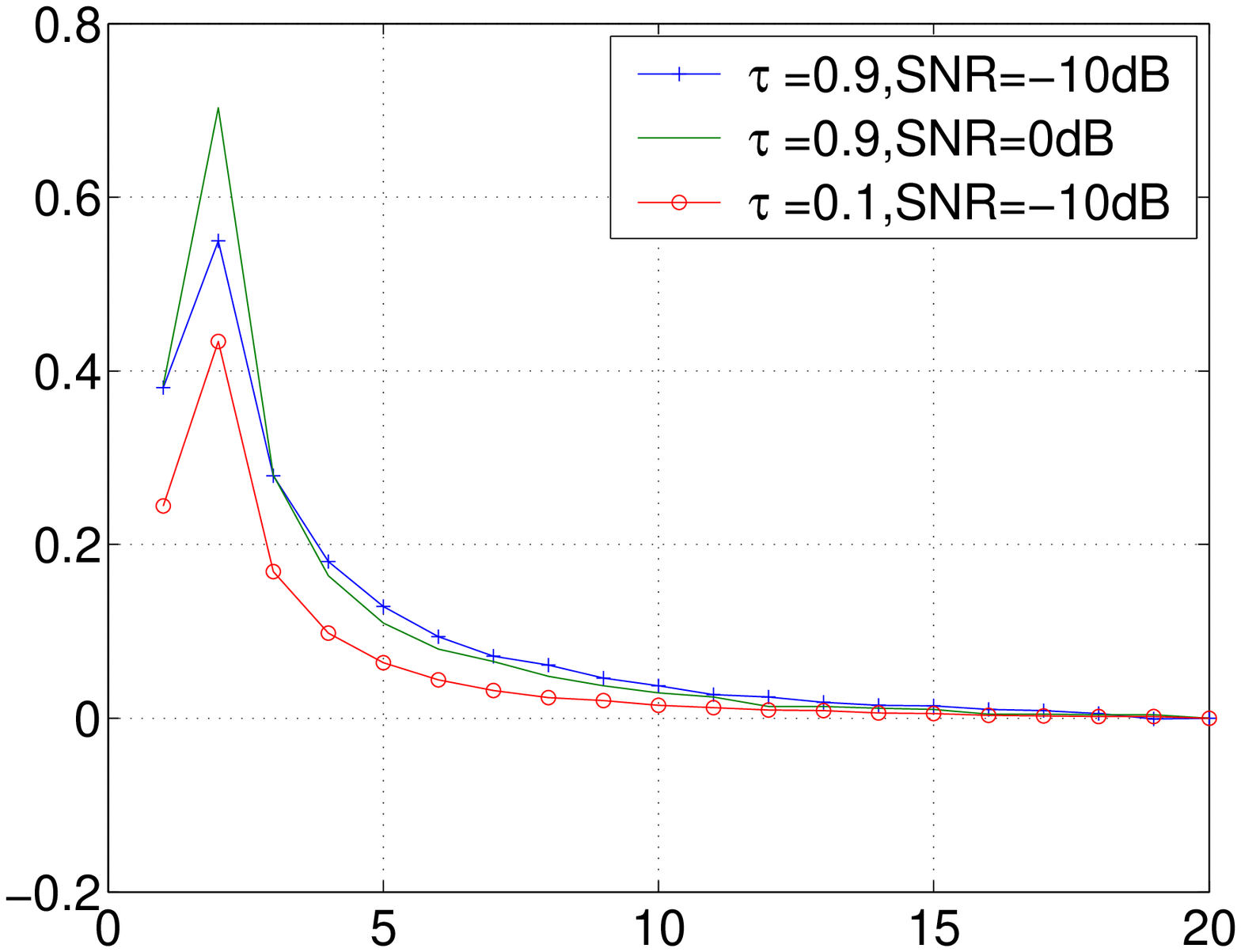}}
  \put(27,1){\footnotesize  iterations}
  \put(0,30){\rotatebox{90}{\footnotesize $C-\Psi(\herm{\trimat}\trimat)$}}
\end{picture}
}
\end{center}
\caption{Converge of Algorithm~\ref{Alg:1}, for various covariance matrices $T=\tau \ones + \diag\{\tau-1\}$ \eqref{E:Tdef:ones} and random rank-one mean, $M_o$. Each plot is averaged over several independent choices  of $M_o$. Figure~\ref{F:conv:F1} shows convergence  for $5\times5$ matrices,  Figure~\ref{F:conv:F2} shows convergence for $15\times15$ matrices and Figure~\ref{F:conv:F3} shows convergence for $25\times25$ matrices }\label{F:conv_nxn}
\end{figure}

In Figure~\ref{F:conv_nxn} we  have shown the convergence of Algorithm~\ref{Alg:1}  for different matrix dimensions, correlations for $T$  and SNR values. In each plot the channel is a non-zero mean, correlated Gaussian, $H\sim\normal{n}{n}{M_o}{I\otimes T}$. Where $M_o = \mu\herm{\mu}$ for a random vector $\mu\in\complex^{1\times n}$. The plots have been averaged over different values of $M_o$. Each convergence is run independently with a random seed value of $\trimat$. Algorithm~\ref{Alg:1} converges to the capacity of the channel, although the convergence rate decreases for larger dimensions. As the channel dimension (and/or SNR) increases, the algorithm becomes more reliant on accurate Monte-Carlo integration, and thus individual iterations take an increasingly long time.

\subsection{Gaussian channel, non-commuting mean and covariance}
Consider a channel where 
\begin{align}
H &= \kappa M_o + (1-\kappa) X \label{E:ex1}
\\
 X &\sim \normal{m}{m}{0}{I\otimes\Sigma}, 0\leq \kappa \leq 1
\end{align}
using the notation of~\cite{GuptaNagar99}. Further, we shall assume that the matrices $M_o$ and $\Sigma$ may not be jointly diagonalized (which is equivalent to the Hermitian matrices $M_o$ and $\Sigma$ being non-commuting~\cite[pp. 229]{Horn90}). We ask: \emph{How does the optimal covariance relate to $M_o$ and $\Sigma$ as $\kappa$ varies between 0 and 1?}

For the purpose of providing graphical results we shall limit ourselves to a $2\times2$ case. While the numerical solution of this problem is straight-forward with Algorithm~\ref{Alg:1}, describing the outcome poses several problems: it is insufficient to investigate only the entries of $\hat{Q}$, since the subspace over which the optimal $Q$ acts will change as $\kappa$ varies.  
%\begin{figure}%[tbp]
%\begin{center}
%\setlength{\unitlength}{0.7mm}
%\begin{picture}(105,75)
%\put(1,0){\includegraphics*[width=\figwidth]{paper_fig2}}
%\put(0,12){\rotatebox{90}{\footnotesize Interpolation factor $\kappa$}}
%\end{picture}
%\caption{Variation in eigenvectors of optimal covariance matrix $Q$, with $H = \kappa X\Sigma^{1/2} +(1-\kappa)M_0$. The eigenvectors of $\Sigma$ are superimposed on the plane $\kappa=0$ and the eigenvectors of $M_o$ are superimposed on the plane $\kappa=1$.
%%
%}\label{F:vect_path_1}
%\end{center}
%\end{figure}

\begin{figure}%[tbp]
\begin{center}
\setlength{\unitlength}{0.7mm}
\begin{picture}(105,75)
\put(1,0){\includegraphics*[width=\figwidth]{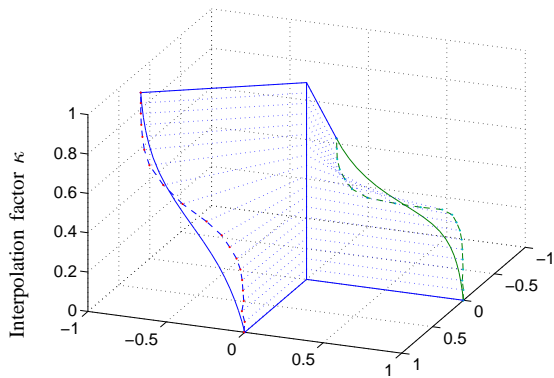}}
\put(0,12){\rotatebox{90}{\footnotesize Interpolation factor $\kappa$}}
\end{picture}
\caption{Variation in eigenvectors of optimal $2\times2$ covariance matrix $Q$, with $H = \kappa X\Sigma^{1/2} +(1-\kappa)M_0$. Eigenvectors of $Q$ are shown dashed. The eigenvectors of $\Sigma$ are superimposed on the plane $\kappa=0$ and the eigenvectors of $M_o$ are superimposed on the plane $\kappa=1$. The eigenvectors of $E\{\herm{H}H\}$ are given as solid lines, superimposed over the dashed lines corresponding to $Q$.
}\label{F:vect_path_1}
\end{center}
\end{figure}

We note that the optimal covariance has eigenvectors which are not trivially related to the eigenvectors of the mean $M_o$ or variance $\Sigma$. Further, the eigenvectors are not given by a direct interpolation between $M_o$ and $\Sigma$, as can be seen by the superimposed the eigenvectors of $E\{S\}$.

Figure~\ref{F:vect_path_1} shows the trajectory of the eigenvectors of
the optimal input covariance $Q=U\hat{Q}U$ as $\kappa$ varies between
0 and 1 for $M_o = \left(\begin{smallmatrix}0 &1\\ 1
    &1\end{smallmatrix}\right)$ and $\Sigma =
\left(\begin{smallmatrix}4 &0\\ 0 &1\end{smallmatrix}\right)$.  The
points are plotted by writing the columns of $U$ as two points in
$\real^2$.  The
vertical axis shows the value of $\kappa$.  On the plane $\kappa=0$, the channel
is zero-mean, correlated Gaussian $H\sim\normal{2}{2}{0}{\Sigma}$. It can be seen that the
power allocation is divided between the eigenvectors of the covariance
matrix $\Sigma$. Similarly, on the plane $\kappa=1$, the channel is
deterministic, with $H=M_o$. The optimal strategy in this case is
beamforming. At each end of the plot, the singular vectors of $M_o$
and $\Sigma$ have been superimposed, for comparison with $Q$.

\begin{figure}
\setlength{\unitlength}{0.7mm}
\begin{center}
\begin{picture}(105,75)
\put(1,0){\includegraphics*[width=\figwidth]{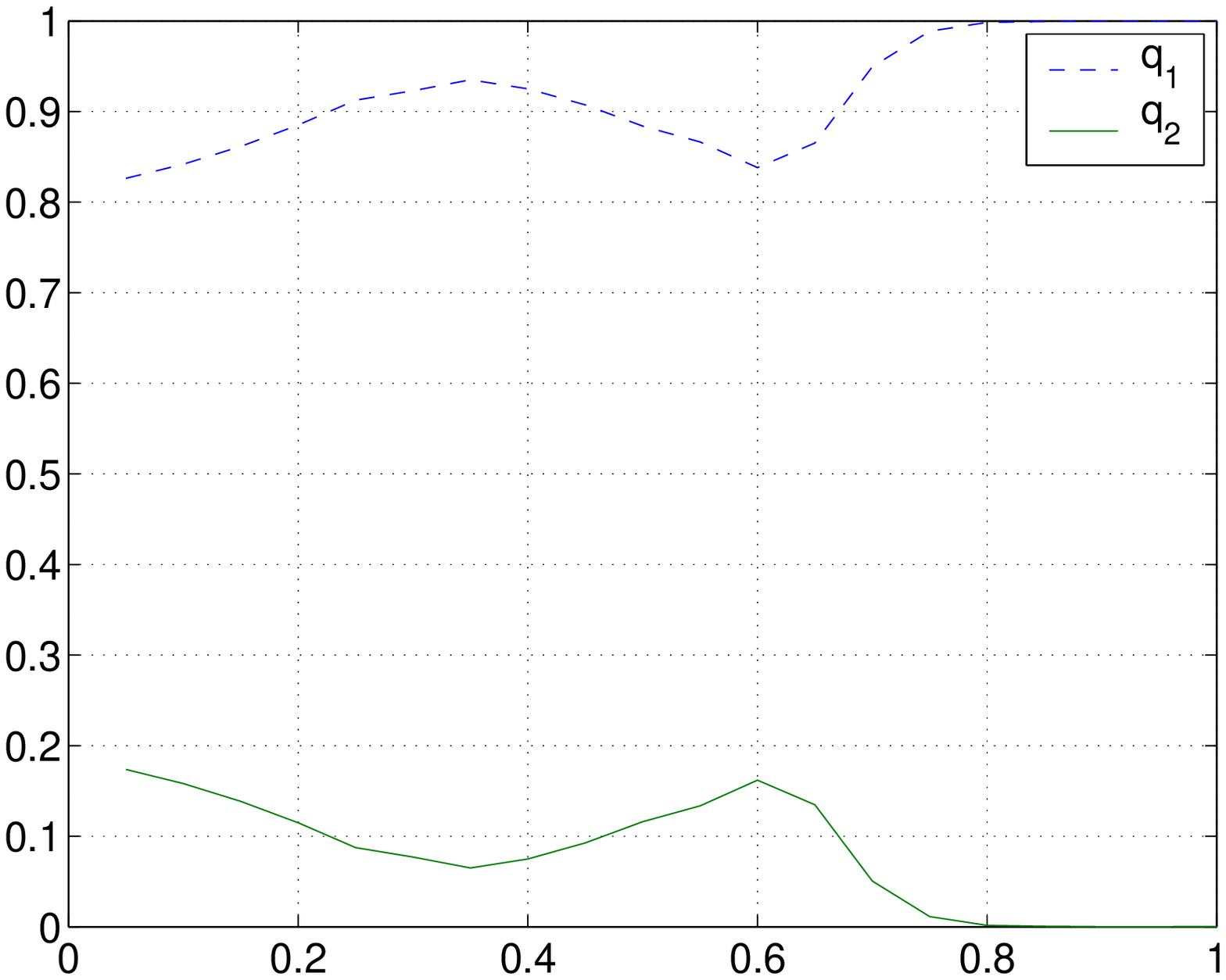}}
\put(0,12){\rotatebox{90}{\footnotesize Power allocation, $q_1,q_2$}}
\put(32,-5){\footnotesize Interpolation factor $\kappa$}
\end{picture}
\end{center}
\caption{Power allocation for optimal input covariance, $\hat{q}_1$ and $\hat{q}_2$ with $Q=U\hat{Q}\herm{U}$}\label{F:power_alloc}
\end{figure}

\subsection{Asymptotics}\label{sec:asymptotics}
It is interesting to consider the low- and high-SNR asymptotics of the
MIMO channel capacity. This has been done by many authors. Here we
give a brief analysis, and in the spirit of the main result presented
above, emphasize the results which hold for any $p_H$.

Consider the matrix channel \eqref{E:1.the_channel} and define
$S= H Q \herm{H}$.  By Taylor series expansion, (\ref{E:def-psi})
may be approximated near $\snr=0$ by
\begin{equation}
  \Psi(Q) \approx \sum_{n=1} (-1)^{n-1}
  \frac{\snr^n}{n}\expect{\tr(S^n)}.\label{E:Taylor_n} 
\end{equation}
where $S=\herm{H}H$.  Of particular interest is the first order
approximation, $\Psi(Q) \approx \snr \tr\left( Q \expect{\herm{H}H}
\right)$.
\begin{theorem}[Low SNR]\label{C:1.1}
  Consider a matrix channel \eqref{E:1.the_channel}, with
  $\expect{H\herm{H}} = U\Lambda\herm{U}$, with $U$ unitary and
  $\Lambda$ diagonal with
  $\Lambda=\diag\{\lambda_1,\ldots,\lambda_\nt\}$ and
  $\lambda_1=\cdots=\lambda_k>\lambda_{k+1}\cdots>\lambda_\nt>0$. For
  low SNR, $\snr\lambda_1\ll1$ the capacity achieving distribution is $Q
  = U\hat{Q}\herm{U}$ where $\hat{Q}$ is diagonal and
\begin{align*}
\hat{Q} = \diag&\left\{ \underbrace{\frac{1}{k},\ldots,\frac{1}{k}},0,\ldots,0\right\}\\
& \quad \text{$k$ terms}
\end{align*}
and $C = \snr k \lambda_1$.
\end{theorem}
At low SNR the transmitter only needs to know $\expect{H\herm{H}}$,
regardless of the underlying $p_H$. To first order, beamforming in the
direction of the largest eigenvector of $\expect{H\herm{H}}$ is
optimal (assuming a unique largest eigenvalue). This aligns with well
known results~\cite{SimMouJSAC03,JafGold:ISIT03}.

This result must be taken with care: the approximation is for $\snr
\lambda_1 \ll 1$ so that large channel gains will necessitate a
correspondingly smaller value of $\snr$ before the expansion is
accurate, see for example~\cite{SimMouJSAC03,JafGold:ISIT03}.

For Ricean channels with separable correlation, a closed form result
may be obtained. Suppose $H \sim \normal{\nt}{\nr}{M}{\sr\otimes\st}$,
where none of $M$, $\sr$ or $\st$ are assumed to be diagonal, or
jointly diagonalizable.  From~\cite[pp.~251]{GuptaNagar99},
$S=H\herm{H}$ is a quadratic normal form and
  \begin{equation}\label{eq:ehh}
    \expect{H\herm{H}} = \st\tr(\sr) + \herm{M}M.
  \end{equation}
  thus
  \begin{equation}
    \left.C(\snr)\right|_{\snr\to0} = \snr \lambda_1
  \end{equation}
  where $\lambda_1$ is the largest eigenvalue of $\st\tr(\sr) +
  \herm{M}M$. This makes it clear that the most fortuitous arrangement
  of $\st$ and $M$ is when they share a common largest
  eigenvector. for $\sr=I$ and $\nr=\nt$, \eqref{eq:ehh} is essentially
  the central Wishart approximation of Lemma \ref{lem:appwish}. This
  is not coincidence, since the central Wishart approximation is
  found by matching the first moment of the density. 
 
  There are several special cases that result in simpler forms for
  $\lambda_1$.
  \begin{enumerate}
  \item In the case of identity transmit covariance $\st=I_\nt$,
    $\lambda_1 = \tr(\sr) + \lambda_1(\herm{M}M)$.
  \item $M = \alpha I$. Then $\lambda_1 = \alpha^2 + \tr(\sr)\lambda_1(\st)$.
  \item Weak LOS component, $\st \tr(\sr) >> \herm{M}M$. Then
    $\lambda_1 = \tr(\sr) \lambda_1(\st) + \epsilon$, where
    $|\epsilon| \leq \lambda_1(\herm{M}M)$. Obviously if $M=0$,
    $\epsilon=0$. 
  \item Strong LOS component, $\herm{M}M >> \st\tr{\sr}$. Then
    $\lambda_1 = \lambda_1(\herm{M}M) + \delta$, where $|\delta| \leq
    \tr(\sr)\lambda_1(\st)$.
  \item For $\nr = \nt = 2$ it is easy to obtain a closed form solution
    for $\lambda_1$.
  \end{enumerate}

  Turning now to the other extreme, for large $z$, $\log(1 + z)
  \rightarrow \log(z)$, and hence at high SNR,
\begin{equation}\label{E:def.hisnr}
  \Psi(Q) \rightarrow \nt \log \snr +\log\det Q + \log\det(\herm{H}H).
\end{equation}
Care must be taken in the definition of ``high'' SNR. The
approximation~\eqref{E:def.hisnr} is only valid when $\snr Q_{ii} \gg
\lambda_{\min}$, ie. the high SNR, is based on high \emph{received}
SNR over all modes, not necessarily high transmit power.

\begin{theorem}[High SNR]\label{Thm:high.snr}
  Consider a matrix channel \eqref{E:1.the_channel} with $H$ a random
  variable, independent of $Q$. Then the capacity achieving
  distribution is $Q = I_\nt/\nt$ and the resulting capacity is
  \begin{equation}\label{E:thm.high_snr}
    C \rightarrow \nt \log\left(\frac{\snr}{\nt}\right) +
    \expect{\log\det(H\herm{H})} 
  \end{equation}
 for any probability density function
$p_H$, provided that $H$ is independent of $Q$.
\end{theorem}
Theorem~\ref{Thm:high.snr} holds  regardless of the
characteristics of the channel. The optimal transmit strategy at high
SNR is equal power, independent white signals. This is not surprising
when it is seen that for \emph{large} received power, the variation in
channel strength is meaningless. From a water-filling perspective, we
have a very deep pool, with tiny pebbles on the bottom: allocation of
power is irrelevant. The channel distribution $p_H$ has no effect on
the optimal transmission strategy, and only affects the resulting
capacity via the $\expect{\log\det(H\herm{H})}$ term. This is
investigated in much more depth in \cite{LozTulVer:ISIT04,LozTulVer-unpub}.

Note also that at high SNR, $\nt \log (P/\nt)$ is asymptotic to the
capacity resulting from transmitting independent data across $\nt$
non-interfering AWGN channels (each channel getting $P/\nt$ of the
available power). The remaining term is either a capacity loss or gain over
this parallel channel scenario, depending on the statistics of the
channel. In the case of Wishart matrices,
\mbox{$H\sim\normal{\nt}{\nr}{0}{\sr\otimes I}$}
\eqref{E:thm.high_snr} has a known closed-form
solution~\cite{KimLap03}. For numerical purposes,
$\expect{\log\det(H\herm{H})}$ may be obtained by Monte-Carlo methods.

\section{Conclusion}\label{sec:Conclusion}
This paper has shown how to correctly compute the capacity of
multiple-input multiple-output channels whose gain matrices are chosen
independently each symbol interval according to a given matrix
density. The optimal input density is Gaussian but is not identically
distributed over time or space except in special cases. 

In the case of full CSI at the transmitter, the optimal power
allocation corresponds to water-pouring in space and time, and is
performed instantaneously, which is an important practical
consideration. At each symbol, the transmitter still performs water
pouring over the channel eigenvalues at that instant, but uses a water
level that results in the long-term average power constraint being
satisfied. In certain circumstances, this yields a considerable gain
in rate, compared to a symbol-wise water-filling, in which the
transmitter uses a water level that enforces a per-symbol power
constraint. The peak-to-average power ratios and entire power
distribution resulting from the use of the optimal space-time
water-filling strategy were also considered. For Rayleigh channels,
the resulting peak-to-average power ratio can be several decibels,
depending upon the average power.

We have investigated the capacity achieving input covariance in the case where the transmitter has statistical CSI. We have presented a method for calculating the optimal input covariance for arbitrary Gaussian vector channels. We have provided an iterative algorithm which converges to the optimal input covariance, by considering the covariance in terms of a Cholesky factorization. We have demonstrated the algorithm on several difficult channels, where the appropriate ``diagonal'' $Q$ input cannot be readily found by inspection. Although the diagonalizing decomposition $Q=U\hat{Q}\herm{U}$ always exists, we have shown that the matrix $U$ may be non-trivially related to the pdf of the channel.

For special cases, the optimal input covariance can be a-priori
diagonalized by inspection -- such as for zero-mean Kronecker
correlated Rayleigh channels. In such cases we gave a simpler fixed
point equation that characterizes the optimal transmit
covariance. This particular characterization reveals a close link
between the optimality condition for deterministic channels (water
filling) and that for ergodic channels.

\appendix
\section*{Proofs}

\begin{proof}[Proof: Theorem~\ref{48831th:capacity}]
The capacity is given by
\begin{equation}
  \label{48831eq:capacity2}
  C = \lim_{N\rightarrow\infty} \sup_{p(x_N)} \frac{1}{N}
  I\left(x_N;y_N \mid H_N\right). 
\end{equation}
For fixed $N$ re-write the entire sequence of
transmissions~(\ref{E:1.the_channel}) as
\begin{equation}
  \label{48831eq:channelN}
  y_N = H_N x_N + z_N.
\end{equation}
For any fixed value of $N$, the optimal density on $x_N$ is obtained
by water-filling on the $Nm$ eigenvalues
$\nu_1,\nu_2,\dots,\nu_{Nm}$ of $W_N=H_N \herm{H_N}$.  Thus
the optimized information rate for given $N$ is given parametrically
by
\begin{align}
  C_N &= \frac{1}{N}\sum_{i:\nu_i^{-1}\leq\xi} \log \xi \nu_i
  \label{48831eq:capacityC2} \\
  P &= \frac{1}{N} \sum_{i:\nu_i^{-1}\leq\xi} \xi - \nu_i^{-1}.
  \label{48831eq:capacityP2} 
\end{align}
Now for a block diagonal matrix such as $W_N$, the $Nm$ eigenvalues
are simply the set of all the eigenvalues of the component diagonal
blocks, in this case the $H[\symboltime]\herm{H}[\symboltime]$. As $N\rightarrow\infty$, the
distribution of the eigenvalues of $W_N$ converges to the eigenvalue
density $p_\Lambda$ associated with $p_H$ and the summations become
expectations with respect to a randomly chosen eigenvalue of
$H\herm{H}$.
\end{proof}

\begin{proof}[Proof: Theorem \ref{48831th:papr}]
  A few observations can be made regarding the distribution of power
  resulting from the optimal transmit strategy.  Firstly, transmit
  power is upper-bounded by $m\xi$, since the instantaneous power
  level on each eigenvector is $\xi-1/\lambda_i$, and $\lambda_i\geq
  0$. The peak-to-average power ratio (PAPR) is therefore $m\xi/\snr$.
  Now from~(\ref{48831eq:capacityP3}),
\begin{align*}
  \frac{\snr}{m} &= \int_{\xi^{-1}}^\infty
  \left(\xi-\frac{1}{\lambda}\right) f(\lambda)\, d\lambda \\
  &\geq \int_{0}^\infty
  \left(\xi-\frac{1}{\lambda}\right) f(\lambda)\, d\lambda \\
  &= \xi -  E\left[\lambda^{-1}\right].
\end{align*}
The inequality is due to the fact that the portion of the integral
from $0$ to $1/\xi$ is non-positive. Therefore $\xi$ is upper-bounded
\begin{equation*}
  \xi \leq \frac{\snr}{m} + E\left[\lambda^{-1}\right].
\end{equation*}
\end{proof}

\begin{proof}[Proof: Theorem \ref{th:monotonic}]
An optimal $Q$ has eigenvalues with satisfy \eqref{E:opt_cov2}, and hence
\begin{align*}
\frac{1}{\nu} \frac{\partial q_k}{\partial\snr} &= \frac{\partial q_k}{\partial\snr}
\left[\expect{\left((\snr\hat{Q})^{-1}+S\right)^{-1}S}\right]_{kk} \\
&= \left[\left((\snr\hat{Q})^{-1}+S\right)^{-1} \snr^{-2}\hat{Q}^{-1} \left((\snr\hat{Q})^{-1}+S\right)^{-1}S \right]_{kk} \\
&= \left[\left(\snr I + \snr^2\hat{Q}S\right)^{-1}\left((\snr\hat{Q}S)^{-1}+I\right)^{-1} \right]_{kk} \\
&= \left[\left((\hat{Q}S)^{-1}+2\snr I + \snr^2 \hat{Q}S \right)^{-1}\right]_{kk} \\
&= \left[A^{-1}\right]_{kk}
\end{align*}
where $A=\herm{A}\geq 0$ (since $S\geq $ and $Q\geq $ are both Hermitian). Now $\det(A) A^{-1} = \adj(A)$ and the diagonal elements of $\adj(A)$ are determinants of principal minors of $A\geq 0$, which are non negative \cite[p. 398]{Horn90}. Noting that $\partial \nu/\partial\snr>0$ completes the proof.
\end{proof}

\begin{proof}[Proof: Theorem~\ref{th:optbeam}]
  Rank-one transmission with $Q=E_{11}$ is optimal if reduction in
  $q_1$ (and corresponding increase in some other $q_i$ results in an
  overall decrease in mutual information. From the Kuhn-Tucker
  conditions \eqref{eq:KT1}, \eqref{eq:KT2}, the condition for
  optimality is (see
  also~\cite{VisMad01,JafVishGold:ICC01,JafGol01ISIT,GolJaf03JSAC,MousSim:IT03})
\begin{equation}\label{eq.p:optbeam}
  \left. \frac{\partial\Psi}{\partial q_1} \right|_{Q=E_{11}} \geq
  \left. \frac{\partial\Psi}{\partial q_k} \right|_{Q=E_{11}} \quad k\geq 2.
\end{equation}
Furthermore, we can restrict attention to $k=2$ in \eqref{eq.p:optbeam}.

Now
\begin{equation*}
  \frac{\partial}{\partial q_k} \Psi(Q) =
  \expect{\left(\left(I+ SQ\right)^{-1} S\right)_{kk}}
\end{equation*}
where $S = \snr \st^{1/2} \herm{X}\sr X\st^{1/2}$ with
$X\sim\normal{\nt}{\nr}{0}{I}$. 

Now $A=I+SE_{11}$ is of the form
\begin{equation*}
  \begin{pmatrix}
    1+S_{11} & 0_{m-1} \\
    b & I_{m-1}
  \end{pmatrix}
\end{equation*}
where $0_{m-1}$ is an all-zero row vector of length $m-1$ and $b$ is a
column vector of length $m-1$. We need to find the inner product
between row $k\geq 2$ of $A^{-1}$ and the corresponding column $k$ of
$S$. Applying the partitioned matrix inverse theorem yields
\begin{equation*}
  A^{-1} =
  \begin{pmatrix}
    \frac{1}{1+S_{11}} & 0_{m-1} \\
    \frac{-b}{1+S_{11}} & I_{m-1}
  \end{pmatrix}
\end{equation*}
and hence for $k>1$,
\begin{align*}
\left. \frac{\partial\Psi}{\partial q_k} \right|_{Q=E_{11}}
&= \expect{S_{kk}} - \expect{\frac{S_{k1}S_{1k}}{1+S_{11}}}  \\
&\stackrel{(a)}= \expect{S_{kk}} - \expect{\frac{|S_{1k}|^2}{1+S_{11}}} \\
&\stackrel{(b)}= \snr \nr \sti_k - \expect{\frac{|S_{1k}|^2}{1+S_{11}}} \\
&\stackrel{(c)}=  \snr \nr \sti_k - \expect{ \frac{\snr^2\sti_1\sti_k
    \left(\sum_{i=1}^\nr \sri_i\, X_{i1}^* X_{ik}\right)^2}{1+\snr\sti_1
    \sum_{i=1}^\nr \sri_i\, |X_{i1}|^2 } }  
\end{align*}
since (a) $S=\herm{S}$, (b) $\expect{S} = \snr\tr(\sr)\st$, and (c),
\begin{equation*}
  S_{1k} = \snr \sqrt{\sti_1\sti_k} \sum_{i=1}^\nr \sri_i\, {X}_{i1}^* X_{ik}.
\end{equation*}
Similarly, for $k=1$
\begin{align*}
  \left. \frac{\partial\Psi}{\partial q_1} \right|_{Q=E_{11}} &= 
\expect{\frac{S_{11}}{1+S_{11}}} \\
&= \expect{ \frac{\snr\sti_1
    \sum_{i=1}^\nr \sri_i\,|X_{i1}|^2}{1+\snr\sti_1
    \sum_{i=1}^\nr \sri_i\, |X_{i1}|^2 } }
\end{align*}
Finally, the expectation with respect to the $X_{ik}$ may be taken, which completes the proof (using the fact that the $X_{ik}$ are independent of the $X_{i1}$).
\end{proof}

\begin{proof}[Proof: Theorem \ref{th:optbeam2}]
  We need to compute the expectation \eqref{eq:optbeam} where $W =
  \herm{X}\sr X$, with $X\sim\normal{\nr}{2}{0}{I}$. To that end, let
  $u\sim\normal{\nr}{1}{0}{I}$ and $v\sim\normal{\nr}{1}{0}{I}$ be
  independent Gaussian random vectors. Then $W_{11} \sim \herm{u}\sr
  u$ and $W_{12} \sim \herm{u}\sr v$.  Noting that $\int_0^\infty
  e^{-xz}dx =1/z$, (which was also a key step for \cite{SimMou03}),
\begin{align*}
  E &= \int_0^\infty e^{-x} \expect{\exp\left({-x\snr\sti_1 \herm{u}\sr u}\right)
    \left(\herm{u}\sr u + \snr\sti_1 |\herm{u}\sr v|^2 \right)} \, dx
  \\
  &= \int_0^\infty e^{-x} \expect[u]{\exp\left({-x\snr\sti_1 \herm{u}\sr u}\right)
    \left(\herm{u}\sr u + \snr\sti_1 \expect[v]{|\herm{u}\sr v|^2}
    \right)} \, dx \\
  &= \int_0^\infty e^{-x} \expect[u]{\exp\left({-x\snr\sti_1
        \herm{u}\sr u}\right) 
    \left(\herm{u}\sr u + \snr\sti_1 \herm{u}\sr^2 u
    \right)} \, dx
\end{align*}
since $u$ and $v$ are independent. Now define $a_i =
\snr\sti_1\sri_i$, let $w_i=|u_i|^2$ (with density $e^{-w_i}$).
Writing out the inner products as summations and using the properties
of the exponential,
\begin{align*}
  E &= \int_0^\infty e^{-x} \expect{\prod_{j=1}^\nr e^{-xa_j w_j}
    \sum_{i=1}^\nr \left(\sri_i + \snr\sti_1\sri_i^2\right) w_i} \, dx
  \\
  &= \int_0^\infty e^{-x} \sum_{i=1}^\nr \left(\sri_i +
    \snr\sti_1\sri_i^2\right) \expect{w_i e^{-xa_i w_i}}\prod_{i\neq
    j} \expect{e^{-xa_j w_j}} \, dx
\end{align*}
where the last line is due to the independence of the $w_i$. Computing
the expectations results in
\begin{align*}
  E &= \sum_{i=1}^\nr \left(\sri_i + \snr\sti_1\sri_i^2\right)
  \int_0^\infty e^{-x} \frac{\sri_i}{(1+a_i x)} \prod_{j}
  \frac{1}{1+a_j x} \, dx \\
  &= \sum_{i=1}^\nr \left(\sri_i + \snr\sti_1\sri_i^2\right)
  \int_0^\infty e^{-x} \frac{\sri_i}{(1+a_i x)} \sum_{j}
  \frac{a_j^{\nr-1}}{1+a_j x} \prod_{k\neq j}(a_j-a_k)^{-1} \, dx
\end{align*}
via partial fraction expansion of the product. Exchanging the order of
integration and summation and noting
\begin{equation*}
  \int_0^\infty \frac{e^{-x}}{(1+a_i x)(1+a_j x)}
  \, dx = \zeta_{ij}
\end{equation*}
as defined in the statement of the theorem completes the proof (with a
few algebraic re-arrangements).
\end{proof}

%This is generated by
%following the deterministic case.
 %
%\begin{equation}
%L(\tau,\mu,\nu) = \Psi(T^*T,S) + \mu\left(\sum_{i\leq j}^t \tau_{ij}^2 \right) - \sum_i\nu_i\tau_{ii}
%\end{equation}
%For this we use the following 
 %

\begin{proof}[Proof: Lemma~\ref{Lem:1}]
  We only consider entries in the upper-triangular (non-zero) part of
  $\trimat$, $\trimati_{i\leq j}$. We need $Q=\herm{\trimat}\trimat$ with $\tr(Q) = \sum_{i<j}(\trimati_{ij})^2 = 1$ and the diagonal elements of $\trimat>0$. We will minimize the negative of $f(\nu)$ 
\begin{itshape}
Minimize $-f(\nu)$ subject to 
\begin{align*}
k_1 &= \sum_{i<j} (\nu_{ij})^2 - 1 \leq 0\\
g_i &= -\nu_{ii} \leq 0
\end{align*}
\end{itshape}  
  
%\begin{align*}
%\mu: &\sum_{i<j} (\nu_{ij})^2 = 1\\
%\phi_k: &\nu_{ii} > 0 \quad k=1,\ldots,\nt
%\end{align*}

Create a modified cost function $J(\nu,\mu,\phi)$ to be minimized, given by
\begin{align*}
J\left[f(\nu),\mu,\phi\right] &= -f(\nu)+\mu k_1(\nu)+\sum_{i=1}^t\phi_i g_i(\nu)\\
&= -f(\nu)+\mu\left(\sum_{i<j}\nu_{ij}^2-1\right)+\sum_{i=1}^t\phi_i(-\nu_{ii})
\end{align*}

We wish to find $\min_\nu J\left[f(\nu),\mu\phi\right]$. The first step is to find the conditions for the optimal point $\nu^o$   to be a minimum.
From~\cite{KuhnTucker51,BeveridgeSchechter70,BoydVanden04}   $\nu^o$   must satisfy
\begin{enumerate}
\item $J\left[f(\nu),\mu,\phi\right]$ is stationary at the optimal point $\nu^o$  
\item $\sum_i \mu_i k_i(\nu^0)=0$ for every  constraint $k_i(\nu)$
\item $\mu_i \geq 0$ $\forall i$.
\item If $\mu_i\neq0$ then constraint $k_i(\nu)=0$ 
\end{enumerate}
From item 1,
\begin{equation}\label{E:p1.stationary_pt}
\frac{\partial L(\nu,\mu,\nu)}{\partial \nu_{ij}} = -\frac{\partial f(\nu)}{\partial \nu_{ij}} +2\mu \nu_{ij} - \phi_i \delta_{ij}=0, \quad \mu,\phi_i\geq0
\end{equation}
where $\delta_{ij}$ is the Kronecker Delta, $\delta_{ij}=1$ for $i=j$. Rearranging \eqref{E:p1.stationary_pt} gives:
\begin{align}
\frac{\partial f(\nu)}{\partial \nu_{ij}} &= 2\mu \nu_{ij}, \quad i\neq j, \mu>0\\
&=2\mu \nu_{ii}, \quad \nu_{ii} > 0, \mu>0\\
&<0\quad \nu_{ii} = 0
\end{align}
%
%\begin{equation}\label{E:p1.partial-diff}
%%\frac{\partial L(\nu,\mu,\nu)}{\partial \nu_{ij}} = 0 =
%\begin{cases}\displaystyle
%\frac{\partial f(\nu)}{\partial \nu_{ij}}
%+2\mu \nu_{ij} & i\neq j 
%\\
%\displaystyle
%\frac{\partial f(\nu)}{\partial \nu_{ii}}
%+2\mu \nu_{ii} - \phi_i & i=j
%\end{cases}
%\end{equation}
 %
%Rearranging the first case of \eqref{E:p1.partial-diff} gives \eqref{E:kt-eq} directly, in the case $i\neq j$. 
% %
%Considering the second case, if $\nu_{ii}=0$ then $\phi_i>0$, giving \eqref{E:kt-eq1} and if $\nu_{ii}>0$ then $\phi_i=0$ which gives  \eqref{E:kt-eq2}.
 %
\end{proof}

\begin{proof}[Proof. Theorem~\ref{Th:1.opto}]
For a channel \eqref{E:1.the_channel} where $H$ is defined by an arbitrary pdf, and the receiver has full knowledge of $H$, whilst the transmitter has statistical knowledge, the input distribution is known to be Gaussian with certain covariance~\cite{Ericson:IT70}. Thus it remains to find the optimal covariance $Q^{\text{opt}}$ of the Gaussian input signal. 

Before applying Lemma~\ref{Lem:1} we must show that $\log\det(I+M\herm{X}X \herm{M})$ is convex $\cap$ on any positive definite matrix $X$ -- which implies $\psifunc{\herm{X}X}{S}$ is convex $\cap$ on any positive triangular matrix as we require. Applying a variation of \cite[pp.466-467]{Horn90}.
\begin{align*}
&\log\det\left(I+M\herm{(\alpha A + (1-\alpha) B)}(\alpha A + (1-\alpha) B)\herm{M}\right)\\
&\geq
\log\det\left(I+\alpha^2M\herm{A}A\herm{M} + (1-\alpha)^2M\herm{B}B\herm{M}\right)
\\
&=\log\det\left(\alpha I+\alpha^2M\herm{A}A\herm{M} + (1-\alpha)I + (1-\alpha)^2M\herm{B}B\herm{M}\right) 
\\
&\geq \alpha\log\det\left( I+\alpha M\herm{A}A\herm{M}\right) + (1-\alpha)\log\det\left(I + (1-\alpha)M\herm{B}B\herm{M}\right)
\end{align*}

The result of Theorem~\ref{Th:1.opto} is given by applying Lemma~\ref{Lem:1} to the (convex $\cap$) function $f(\trimati)=\psifunc{Q=\herm{\trimat}\trimat}{S}$. The matrix $Q$ may now be full, but remains positive semi-definite. 
Substituting $X(\trimati)=\herm{\trimat}\trimat$
\begin{align*}
\frac{\partial \psifunc{\herm{\trimat}\trimat}{S}} {\partial \trimati_{ij}} =
%
%\frac{\partial \expect[S]{\log\det(I+SX)}}{\partial \trimati_{ij}} \\
&= \expect[S]{\tr\left[\frac{\partial \log\det\left(I+S X \right)}
{\partial X} \frac{\partial{X} }{\partial \trimati_{ij}}\right]}
\\
%&= \expect[S]{\tr\left[ (I+SX)^{-1} \frac{\partial (I+SX)}{\partial X} \frac{\partial{X}} {\partial \trimati_{ij}}\right]}
%\\
&= \expect[S]{\tr\left[ (I+SX)^{-1} S \frac{\partial{X}} {\partial \trimati_{ij}}\right]}\\
&= %
\expect[S]{\trace\left[(I+S\herm{\trimat}\trimat)^{-1}S\cdot\frac{\partial \herm{\trimat}\trimat}{\partial \trimati_{ij}}\right]}
\end{align*}
Since $\trimati_{ij}$ and $S$ are independent the trace, expectation and differentiation all commute, and the second line arises from application of the matrix chain rule. Observe that $\partial f(X(t))/\partial t = \tr(\partial f(X)/\partial x \cdot \partial X/\partial t) $.
Define $E^{ij}$ as the matrix of partial derivatives of $\herm{\trimat}\trimat$ with respect to $\trimati_{ij}$. In general this matrix is full.
\begin{equation*}
E^{ij} = \frac{\partial (\herm{\trimat}\trimat)}{ \partial \trimati_{ij} }= \frac{\partial \sum_k \overline{\trimati_{mk}} \trimati_{nj}}{\partial  \trimati_{ij}}
\end{equation*}

%The remainder of the the theorem follows by substituting $X=\herm{T}T$ and $z_{ij} = t_{ij}$.
The channel capacity is also known to be the expectation of $\psifuncrand{Q=\herm{\trimat}\trimat}{S=\herm{H}H}$ over $S$, with Gaussian input~\cite{Ericson:IT70}.

\end{proof}

\begin{proof}[Proof: Theorem~\ref{Prop:1}]
  The algorithm is a gradient descent algorithm on a convex
  problem.
\end{proof}

\begin{proof}[Proof: Theorem~\ref{C:1.1}] 
  The optimization may may be written as
\begin{equation}\label{E:pl1}
C = \max_{\tr(Q)=1}\sum_{i=1}^\nt\expect[H]{\log(1+\snr\alpha_i)}
\end{equation}
where $\alpha_i$ is the $i^{th}$ largest singular value of $S=H Q \herm{H}$. Taylor expansion of \eqref{E:pl1}, around $\snr=0$ gives:
\begin{equation*}%\label{E:pl2}
C = \max_{\tr(Q)=1}\sum_{i=1}^\nt\expect[H]{\snr\alpha_i}
= \max_{\tr(Q)=1}\snr\expect[H]{\tr\left(H Q \herm{H}\right)}
\end{equation*}
 %
%which gives the result of \eqref{E:lem1.c}. 
It now remains to find the capacity achieving distribution. 
Note, for any Hermitian matrices $A$ and $B$ with eigenvalues $a_1\geq \cdots\geq a_n$ and $b_1\geq \cdots\geq b_n$, %
\begin{equation*}
\tr(AB)\leq\sum_i a_i b_i
\end{equation*}
with equality if $A$ and $B$ are jointly diagonalizable~\cite{Muirhead82}\footnote{$A=U\hat{A}\herm{U}$ and $B=U\hat{B}\herm{U}$ for diagonal $\hat{A}$ and $\hat{B}$}. With $A=Q$ and $B=\expect{H\herm{H}}$ the capacity achieving distribution diagonalizes $\expect{H\herm{H}}$.
Apply Definition~\ref{D:lagrange} to give %
\begin{equation*}
\left.\frac{\partial I(Q,\snr)}{\partial \hat{Q}_{ii}}=\lambda_i=\mu\right|_{Q_{ii}>0}
\end{equation*} 
Since we require $\mu$ constant for all non-zero $Q_{ii}$, the only valid solution is
\begin{equation*}
Q_{ii} = \begin{cases}1 &i=1\\0 &\text{else}\end{cases}
\end{equation*}
for distinct $\lambda_i$, which gives %\eqref{E:lem1.beamform} 
and substituting for \eqref{E:pl1} gives the desired result.
%\end{proof}

%\begin{proof}[Corollary~\ref{C:1.1}] 
% In the case of $k$ equal values, $$\lambda_1=\cdots=\lambda_k>\lambda_{k+1}\geq\cdots$$ the 
For $k$ equal eigenvalues the unique solution becomes
%\begin{equation*}
%Q_{ii} = \begin{cases}\frac{1}{k} &i\leq k\\0 &\text{else}\end{cases}
%\end{equation*}
$\mu=1/k$, which gives the desired result.
\end{proof}

\begin{proof}[Proof: Theorem~\ref{Thm:high.snr}]
Starting from the definition of high-SNR, note that $I(Q,\snr)$ is dependent on $Q$ \emph{only} through the eigenvalues of $Q$, and not through any interaction with $H$. Using a Lagrange-multiplier method, and differentiating \eqref{E:def.hisnr} with respect to $Q_{ii}$, gives:
\begin{equation*}
\frac{1}{Q_{ii}}=\mu \quad Q_{ii}>0
\end{equation*}
with the only solution, 
\begin{equation*}
Q_{ii} = \frac{1}{\mu} = \frac{1}{\nt}
\end{equation*}
Substituting in \eqref{E:def.hisnr} gives \eqref{E:thm.high_snr}.
\end{proof}

\end{document}